\documentclass[preprint,aps]{revtex4}

\usepackage{tabularx}
\usepackage{bm}
\usepackage{euscript}
\usepackage{epsfig,psfrag,subfigure}
\usepackage{graphicx}
\usepackage{color}
\usepackage{amsfonts}
\usepackage{exscale}
\usepackage{amsbsy}
\usepackage{stmaryrd}
\usepackage{wasysym}

\def\be{\begin{equation}}       \def\ee{\end{equation}}
\def\bea{\begin{eqnarray}}      \def\eea{\end{eqnarray}}
\def\ba{\begin{array} }
\def\ea{\end{array} }
\def\bnum{\begin{enumerate} }
\def\enum{\end{enumerate}}

\def\=>{\Rightarrow}
\def\>{\rightarrow}

\def\eye2{Fathbb{I}}

\begin{document}

%\centerline{ \bf
\title{\bf
Quenched disorder and vestigial  nematicity in the pseudo-gap regime of the cuprates}
\author{Laimei Nie$^1$, Gilles Tarjus$^2$ and S. A. Kivelson$^1$}
\affiliation{1) Department of Physics, Stanford University, Stanford, California 94305, USA}
\affiliation{2) LPTMC, CNRS-UMR 7600, Universit\'e Pierre et Marie Curie, 4 Place Jussieu, 75252 Paris cedex 05, France}
\begin{abstract}
%Incommensurate charge-density-wave short-range order (i.e. with a finite correlation length) and a sharp phase transition to a phase with long-ranged  nematic (possibly chiral-nematic) order is shown to be natural in the presence of weak quenched disorder in systems which, in the absence of disorder, would have unidirectional (stripe) ordered ground states.  Recent experiments probing charge order in the pseudo-gap regime of the hole-doped cuprate high-temperature superconductors are interpreted in light of these results.
The cuprate high-temperature superconductors have been the focus of unprecedentedly intense and sustained study not only because they are, by a wide margin, the materials with the highest superconducting transition temperatures, but also because they represent the most exquisitely studied examples of  highly correlated electronic materials.  In particular, the pseudo-gap regime of the phase diagram, which is the ``normal phase'' out of which the superconductivity, for the most part, develops, exhibits a variety of mysterious emergent behaviors.  In the last few years, evidence from NMR/NQR\cite{NMRYBCO,julien2013} and STM\cite{parker2010,mesaros2011,neto2013,comin2013} studies, as well as from a new generation of X-ray scattering experiments\cite{XrayYBCO,HaydenYBCO,blackburn2012,keimer2013,RexsYBCO} has accumulated indicating that a general tendency to short-range-correlated incommensurate charge-density-wave (CDW) order is ``intertwined''\cite{fradkin2012} with the superconductivity in the pseudo-gap regime.  Additionally, transport\cite{ando2002,daou2010}, STM\cite{rmp,lawler2010}, neutron-scattering\cite{hinkov2008}, and optical\cite{xia2008,he2011,karapetyan2012,karapetyan2013} experiments have produced evidence -- not yet entirely understood -- of the existence of an associated pattern of long-range-ordered point-group symmetry breaking with an electron-nematic and possibly a chiral-nematic (gyrotropic) character\cite{hosur2013}. We have carried out a theoretical analysis of the Landau-Ginzburg-Wilson effective field theory of a classical incommensurate CDW in the presence of weak quenched disorder. While the possibility a sharp phase transition and long-range CDW order are precluded in such systems, we show that any discrete symmetry breaking aspect of the charge order -- nematicity in the case of the unidirectional (stripe) CDW we consider explicitly -- generically survives up to a non-zero critical disorder strength.  Such ``vestigial order,'' which is subject to unambiguous macroscopic detection, can serve as an avatar of what would be CDW order in the ideal, zero disorder limit. Various recent experiments in the pseudo-gap regime of the hole-doped cuprate high-temperature superconductors are readily interpreted in light of these results.

\end{abstract}
\date{\today}
%\date{}                                           % Activate to display a given date or no date
\maketitle
Because the spontaneous breaking of a continuous symmetry is forbidden\cite{imryma} in the presence of ``random-field'' disorder in dimension $d \leq 4$, effects of disorder are significant for the physics of incommensurate CDW ordering, even in crystalline materials, such as the high-temperature superconductor YBCO, which can in other respects be considered extremely well ordered. However, because $d=2$ is the lower critical dimension for breaking a discrete symmetry, as in the random field Ising model\cite{imryma,aizeman1989,natterman}, %if, in addition to breaking translational symmetry,
if a putative CDW ground state breaks a discrete symmetry (e.g. a point-group symmetry), a finite temperature transition at which this symmetry is broken will  persist in the presence of weak disorder in $d=3$.

%In this letter
Here, with the case of the cuprates in mind, we study a model of a layered system with tetragonal symmetry which in the absence of disorder undergoes a transition to a unidirectional incommensurate CDW (stripe ordered) phase.  We thus express the density at position $\vec r$ in plane $m$ as
\be
\rho(\vec r,m) = \bar \rho +\left[\psi_x(\vec r,m) e^{iQ x} +\psi_y(\vec r,m) e^{iQ y} +{\rm H.C.}\right] +\ldots
\ee
where $Q$ is the magnitude of the CDW ordering vector, $\psi_{\alpha}$ (with $\alpha=x,y)$ are the two components of a slowly varying complex vector field, and the ellipsis refers to higher harmonics. Broken symmetries are  defined, as usual, by taking the asymptotic long-distance limit of the appropriate thermal ($<\ >$) and configuration averaged ($\overline{\ ; \ }$)  two-point correlation function, $\lim_{|\vec R|\to \infty}\overline{<O^\dagger(\vec r+\vec R, m)O(\vec r, m)>}\equiv \overline{|<O(m)>|^2}$:  In a stripe ordered state, $\overline{<\psi_x>}\neq 0$ and $\overline{<\psi_y>}=0$ (or vice-versa) and ${\cal N} \equiv \overline{<|\psi_x|^2>}-\overline{<|\psi_y|^2>} \neq 0$; in a checkerboard state $\overline{<\psi_x>}=\overline{<\psi_y>}\neq 0$, and ${\cal N}=0$, while in an ``Ising nematic'' phase, $\overline{<\psi_x>}= 0$, $\overline{<\psi_y>}= 0$, and ${\cal N}\neq 0$.  For each of these states, the pattern of broken symmetry could, depending on details of the interactions between neighboring planes, propagate from plane to plane in different ways, thus breaking the point-group symmetries as well as translation symmetry in the $z$ ($\perp$ to the plane) direction in different ways.

In Eq. \ref{calF}, below, we introduce an effective Landau-Ginzburg-Wilson effective field theory expressed in terms of these fields.  For simplicity, we will assume that the interplane couplings (of magnitude $V_z$) are weak compared to the in-plane interactions, and favor identical ordering in neighboring planes; however, %as we will discuss  below, at the end of the paper,
 it is straightforward  to generalize this to cases in which more complex patterns of interplane ordering are favored.  %We will also chose parameters\cite{robertson} such that in the absence of quenched randomness the ground state is stripe ordered. This
The stripe state breaks a continuous ($U(1)=SO(2)$) symmetry (translations) and a discrete $Z_2$  symmetry associated with the choice of whether the stripes are modulated in the $x$ or $y$ direction.  In the nematic phase, translational symmetry is preserved but the point group ($Z_2$) symmetry is still broken.  We obtain explicit results for the phase diagram and various correlation functions of this model using a saddle-point (mean-field) approximation and the replica trick.  %in certain regions of the phase diagram
  For a generalization of the model in which $\psi_{\alpha}$ is taken to be an $N$ component field, and the $SO(2)\times SO(2)\times Z_2$ symmetry of the original model is generalized to $SO(N)\times SO(N)\times Z_2$, this approximation becomes exact in the $N\to\infty$ limit.  %The quenched disorder is treated using the replica trick.
  We also outline a procedure (explored in more detail in the Supplemental Material) to establish a precise correspondence between the effective field theory for the nematic order parameter and a random-field Ising model (RFIM). % combination of a low temperature expansion and a self-consistent phonon approximation;

{\bf Principal Results:}
Because the general behavior of the system can be motivated largely from symmetry considerations starting directly from the assumption of a stripe ordered state, we begin by presenting our key results on the basis of qualitative arguments, and will then discuss how these results follow from the systematic analysis of the effective field theory.

The structure of the phase diagram in the temperature ($T$) and disorder ($\sigma$) plane is shown in Fig. 1.  In the absence of disorder, stripe order necessarily survives up to a non-zero critical temperature, $T_{str}$.  Here, it is possible that there is a single transition to a fully symmetric state, or it is possible, as shown, for the symmetry to be restored in a sequence of two transitions resulting in the existence of an intermediate nematic phase for $T_{str} \leq T< T_{nem}$.
\footnote{Both scenarios occur in the model defined in Eq.\ref{calF}, depending on the value of $V_z$;  the situation shown in Fig. 1 pertains to the case $V_z < 0.37J$, while for larger $V_z$, there is apparently a single first order transition without an intermediate nematic phase.}
 Assuming the transitions to be continuous, the transitions at $T_{str}$ and $T_{nem}$ are in the 3d-XY and 3d-Ising universality classes, respectively.\footnote{Note that both $T_{str}$ and $T_{nem}$ remain non-vanishing in the limit $V_z\to 0$, although for $V_z=0$ the stripe state would only have quasi-long-range order.}

\begin{center}
\includegraphics[width=10cm]{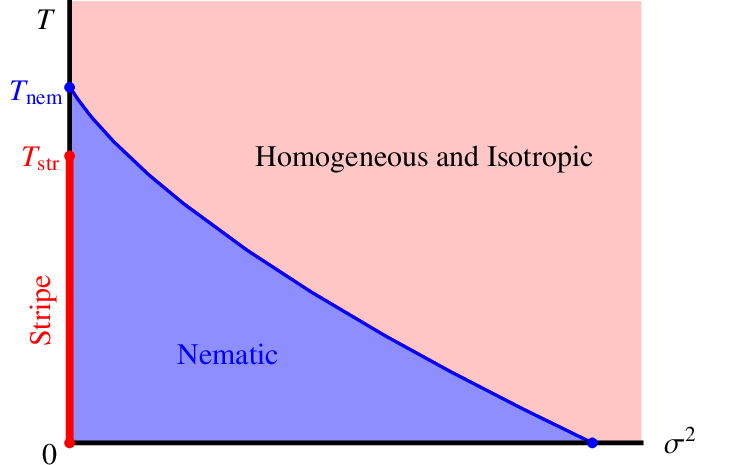}
\makeatletter\def\@captype{figure}\makeatother
\caption{Schematic phase diagram of a highly anisotropic (quasi-2D) tetragonal system as a function of the mean-squared disorder, $\sigma^2$.  The phase diagram is computed from the solution of the self-consistency equations for the lattice version of the model defined in the Supplemental Material with $V_z = 0.01\kappa_{\parallel}, \kappa_{\perp} = 0.98\kappa_{\parallel}$ and $\Delta=0.25\kappa_\|$.}
\end{center}

Non-zero disorder precludes the existence of long-range stripe order;  under some circumstances, for weak enough disorder, the stripe order could give way to quasi-long-range stripe-glass order\cite{giamarchi1994,gingras1996,stripeglass}, but this is not generic\cite{tarjus2006}, %, so for simplicity in the figure we have assumed no such phase occurs.
and is not seen in our effective field theory, at least at the level of the approximate solution we have obtained.  However, the nematic phase has Ising symmetry so it survives as long as the disorder is less than a critical strength, $\sigma_c$.    This is an example of a more general phenomenon, which we have named ``vestigial order''; while the tendency toward stripe order is the essential piece of microscopic physics, the nematic phase is more robust as a phase of matter, and can serve as an avatar of stripe order which can be detected in macroscopic measurements\cite{hosur2013}.  While $\sigma_c$ necessarily vanishes as $V_z \to 0$,  it does so\cite{zachar2003} only as $\sigma_c \sim [\ \log|T_{nem}/V_z|\ ]^{-1/2}$, so it is not too small even in quasi 2D systems.

To obtain explicit expressions for measurable quantities requires making approximations.  One important quantity is the structure factor,  $S(\vec q)$,  which determines the X-ray scattering cross-section.  For $T>T_{str}$, invoking the fluctuation dissipation theorem and linear response analysis, it is straightforward to obtain expressions for $S$ in terms of the susceptibility, $G$, of the ideal (disorder-free) system to second order in $\sigma$.  Specifically,  near the fundamental ordering vectors ($k_x^2+k_y^2 \ll Q^2$),
\bea
S(Q+k_x,k_y,k_z) = &&TG(k_x,k_y,k_z;\mu+ {\cal N})+\sigma^2|G(k_x,k_y,k_z;\mu+{\cal N})|^2\nonumber \\
S(k_x,Q+k_y,k_z) = &&TG(k_y,k_x,k_z;\mu-{\cal N})+\sigma^2|G(k_y,k_x,k_z;\mu-{\cal N})|^2.
\label{structure}
\eea
Even in the limit of weak disorder, this expression is invalid for $T<T_{str}$, reflecting the non-perturbative destruction of long-range CDW order by quenched randomness.  However,
in the Gaussian approximation we define below, which is exact in  the previously mentioned large $N$ limit, an expression of precisely this same form is obtained which is valid for all $\sigma$ and $T$, however with $G$ replaced by an effective susceptibility,
%where in terms of a z-direction dispersion, $\epsilon_z(k_z)$, that depends on the details of the interplane interactions
%({\it e.g.} in the model considered below, $\epsilon_\perp(k_z) = \sin^2(k_z c/2)$),
\be
G(\vec k;\mu) = \left[\kappa_\| k_x^2 + \kappa_\perp k_y^2 +V_z \epsilon_z(k_z) + \mu\right]^{-1} \ .
\label{greenfunction}
\ee
Here  $\epsilon_z(k_z)$ is  the  z-direction dispersion that depends on the details of the interplane interactions, and
 ${\cal N}$  and %the ``mass,''
 $\mu$ are effective couplings which are implicit functions of $T$ and $\sigma$   determined by the self-consistency Eqs. (\ref{hardmu}) and (\ref{calN}), below.
%Since for $\sigma \neq 0$, the CDW order never condenses, it is reasonable to treat its fluctuations in Gaussian %(self-consistent phonon)
%approximation.  In this case, the density structure factor (measurable by X-ray defraction) for in-plane wave-vectors in the neighborhood of the ordering vectors ($k_x^2+k_y^2 \ll Q^2$) is
%$\Delta$ is a coupling constant which enters the effective action;
 In the isotropic phase ${\cal N}=0$, while  in the nematic phase or in the presence of explicit orthorhombic symmetry breaking by the lattice, ${\cal N} \neq 0$.  %The temperature dependence of  ${\cal N}$  and the ``mass,'' $\mu$, are  approximately determined by the self-consistency Eqs. \ref{hardmu} and \ref{calN}, below.
Since there is an actual thermodynamic phase transition involved, direct probes of the nematic phase should, in principle, be possible and unambiguous.  %(Given that nematicity is unnatural in a failed checkerboard state, evidence of nematicity would constitute compelling evidence of an underlying stripe phase.)
There are, however, two aspects of the problem that make this less straightforward than it at first seems.  In the first place, the number of degrees of freedom per unit cell involved in a nematic transition may be relatively small.  For instance, nematic order does not generically open gaps anywhere on the Fermi surface leading to a relatively weak signature in the specific heat\cite{rost2011}, even when the nematic transition occurs at low $T$;  when the transition occurs at relatively high $T$, the smallness of the thermodynamic signal is still more of an issue.  More importantly, since the transition is in the universality class of the %random-field-Ising model (
 RFIM, the intrinsic slow dynamics\cite{natterman} imply that, starting at a cooling-rate dependent temperature strictly larger than $T_{nem}$, the nematic ordering can no longer equilibrate and hence all thermodynamic signatures will be  dynamically rounded.  This is further exacerbated by the fact that any uniaxial strain will couple linearly to the nematic order parameter, so uniaxial strain (or any weak orthorhombicity of the host crystal) will round the transition and random strains will broaden it.

There are, however, clear ways to detect nematic order.  While this has been undertaken in various ways in the context of the cuprates\cite{ando2002,howald2003,hinkov2008,daou2010,lawler2010,parker2010,mesaros2011,julien2013,rmp,vojta2009,pasupathy}, the most successful strategy has been developed in context of studies of the Fe-based high-temperature superconductors.\cite{kuo2013} Several general observations underlie these strategies:

\noindent{\bf 1)}  Any  quantity that is odd under $C_4$ rotations (or the corresponding element of the point group symmetry that is broken in the nematic phase) vanishes in the isotropic phase and grows linearly in proportion to ${\cal N}$ for small ${\cal N}$, and can thus be used as a proxy for the nematic order parameter.  Examples include the resistivity anisotropy\cite{ando2002,borzi2007,chu2010}, $\rho_{xx}-\rho_{yy}$, %the difference in the average of any appropriately defined density on x-directed and y-directed bonds\cite{davisnematic},
any local density which is odd under $C_4$ rotation, or a structural ({\it e.g.} orthorhombic)  distortion\cite{stingl2011}.
Consider, for instance,  the  bond-charge-density on  x-directed and y-directed bonds from site $\vec R$, $\rho_x(\vec R)$ and $\rho_y(\vec R)$, which for the cuprates\cite{julien2013} also corresponds to the charge density on the corresponding in-plane O sites. A direct measure of nematicity is $Q_{\cal N}\equiv \overline{<\rho_x(\vec R)>}- \overline{<\rho_y(\vec R)>} \propto \ {\cal N}$.  A  different measure, which is directly related to local CDW order,   is $\tilde Q_{\cal N} \equiv  \overline{<|\delta\rho_x(\vec R)|^2>}- \overline{<|\delta\rho_y(\vec R)|^2>}$,
where  %  (which vanish identically in the absence of either long-range stripe order or quenched randomness)
\be
\delta\rho_{\alpha}\equiv \rho_{\alpha}-\overline{<\rho_{\alpha}>};\ \  \overline{<|\delta\rho_x(\vec R)|^2>}  = \sigma^2 A_2(\mu+{\cal N});\ \ \overline{<|\delta\rho_y(\vec R)|^2> }=\sigma^2A_2(\mu-{\cal N})
\label{deltarho}
\ee
and
\be
A_p(\mu)\equiv  \int \frac {d\vec k}{(2\pi)^3} \big[G(\vec k,\mu)\big]^p = -\left(\frac 1 {p-1}\right)\frac {\partial A_{p-1}} {\partial \mu}
\label{Am}
\ee
with $G(\vec k;\mu)$ given in Eq. (\ref{greenfunction}). A quantity similar to $Q_{\cal N}$, referred to in Ref. \cite{lawler2010}  as ``intra-unit-cell-nematic'' order, has been investigated in STM studies of cuprate high-temperature superconductors with suitable surfaces.  As has been shown   in Ref. \cite{julien2013} (and discussed below),  bulk  NMR/NQR measurements on cuprates can be performed to obtain $Q_{\cal N}$ and  $\tilde Q_{\cal N}$. % since the average O quadrapolar line-shift is proportional to $\overline{\rho_a(\vec R)}$   and the inhomogeneous quadrapolar broadening  is proportional to $\overline{|\rho_a(\vec R)|^2 }$.

\noindent{\bf 2)} Uniaxial volume preserving strain, $b^{eff}\equiv \epsilon_{xx}-\epsilon_{yy}$,  acts as a symmetry-breaking field (%which is represented by $\vec b^\dagger$ in Eq. \ref{calF}
see Eq. (\ref{beff})) conjugate to ${\cal N}$.  Thus, from the strain dependence of any of the electronic proxies for ${\cal N}$, it is possible to infer the differential susceptibility, $\chi \equiv \partial {\cal N}/\partial b^{eff}$.  Less obviously, but equally importantly, the ability to apply a symmetry-breaking field can, under appropriate circumstances, permit at least two real-world complications to be circumvented:  a)  In an  orthorhombic crystal, there is an explicit symmetry-breaking field which rounds the nematic transition and implies the existence of a non-zero ${\cal N}$ even for $T>T_{nem}$;  however, if the orthorhombicity is sufficiently weak, it is possible\cite{kuo2013} to measure $\chi$ at non-zero $b^{eff}$ and to extrapolate the result to $b^{eff}=0$, thus correcting for the presence of orthorhombicity.  b) Where macroscopic detection of symmetry breaking is precluded  due to domain formation, cooling in the presence of a symmetry breaking field can orient the  order parameter macroscopically, permitting macroscopic measurements to detect its presence.

%\noindent{\bf 3)}  As mentioned previously, the fact that the nematic ordering problem is mapped to an effective RFIM implies slow dynamics which complicate the application of traditional thermodynamic probes of a phase transition.  However, conversely, it implies the existence\cite{carlson} of detectable hysteresis and noise, especially in mesoscopic samples, some of which may have been seen already in mesoscopic samples of YBCO.\cite{vanharlingen}

{\bf Explicit Model:}  To make the present considerations concrete, we consider the simplest classical effective field theory\cite{mcmillan1975,robertson2006,maestro2006} of an incommensurate CDW in a tetragonal crystal, with effective Hamiltonian
\bea
{\cal H} =&& \frac {\kappa_\|} 2 \big| \partial_\alpha \psi_\alpha\big|^2+\frac {\kappa_\perp} 2 \big|\partial_{\bar\alpha}\psi_\alpha\big|^2
+ \frac U {2N}\Big[ \big| \psi_x\big|^2+ \big| \psi_y\big|^2-{\Lambda}N\Big]^2-\frac \Delta {2N}\Big[ \big|\psi_{x}\big|^2-\big|\psi_{y}\big|^2\Big]^2\nonumber \\
&&-V_z\big[ \psi_\alpha^\dagger(\vec r,m) \psi_\alpha(\vec r,m+1)+{\rm H.C.}\big] - \big[ h_\alpha^\dagger(\vec r,m)\psi_\alpha(\vec r,m)+{\rm H.C.}\big] \nonumber \\
&&- \big[ b_\alpha^\dagger\psi_\alpha+{\rm H.C.}\big] +\ldots
\label{calF}
\eea
Here $\alpha=x$, $y$ is a spatial index for which Einstein summation convention  is adopted, $\bar\alpha$ signifies the complement of $\alpha$, and each $\psi_\alpha$  is a $SO(N)$ vector,  where  in the case of the CDW, $N=2$ with the two components corresponding to the real and imaginary parts of the amplitude -- the generalization to arbitrary $N$ permits a controlled solution in the large $N$ limit.  %As usual, we assume that
In the following analysis, we assume that $\Delta >0$, which is to say that stripe order is favored over checkerboard.
In the absence of disorder and significant thermal fluctuations, one might focus on temperatures in the neighborhood of  the mean-field CDW transition temperature, $T_{MF}$, where
$\Lambda<0$ for $T>T_{MF}$ and $\Lambda>0$ for $T<T_{MF}$.
% and other couplings will be assumed temperature independent.
Here, we will focus on the range of temperatures for which $\Lambda>0$, where there is a well developed local amplitude of the CDW order parameter, but in which the effects of weak random fields spoil the long-range CDW ordering at long distances.   We further assume that all the remaining coupling constants are positive. Finally, $h$ is a Gaussian random field,
\be
\overline{h_{\alpha i}(\vec r,m)}=0;\ \ \ \overline{h_{\alpha i}(\vec r,m)h_{\beta j}(\vec r^\prime,m^\prime)} = \sigma^2\delta_{\alpha \beta}\delta_{ij}\delta_{m,m^\prime}\delta(\vec r-\vec r^\prime),
\ee
with $i,j = 1, \ldots, N$, and $b$ is an explicit symmetry-breaking field, which will be assumed to vanish unless otherwise stated.
%In the following analysis, we assume that $\Delta >0$, which is to say that stripe order is favored over checkerboard.  %, and for simplicity we will take the hard spin limit, $U\to \infty$.
 The ellipsis represents higher order terms in the usual Landau-Ginzburg expansion.

 It is convenient to introduce two  scalar Hubbard-Stratonovich fields, $\zeta(\vec r,m)$ and $\phi(\vec r,m)$ in place of the quartic terms in ${\cal H}$:
 \bea
&& \frac U {2N} \Big[\big|\psi\big|^2  -N\Lambda\Big]^2 - \frac \Delta N \big( \big|\psi_{x}\big|^2-\big|\psi_{y}\big|^2\big)^2 \\
\label{calF2}
 &&\ \ \ \ \ \ \to   \frac {\zeta^2} {2U} + \frac {\phi^2}{2\Delta} +\frac {1} {\sqrt{N}}\Big[
i\zeta \big(\big|\psi\big|^2 -\Lambda N\big) + \phi\big( \big|\psi_{x}\big|^2-\big|\psi_{y}\big|^2\big)\Big],
  \nonumber
  \eea
  where $\big|\psi\big|^2 = \big|\psi_x\big|^2 + \big|\psi_y\big|^2$, and in the ``hard-spin limit'' ($U\to\infty$%and $r/U \to -2$
  ) $\zeta$ enforces the hard-spin constraint, $\big| \psi\big|^2=\Lambda N$, and
 % the expectation value of
  $\phi$ determines the nematic order parameter, ${\cal N}=2\langle \phi\rangle/\sqrt{N}$.

{\bf  Approximate Solution: } There are  a number of approximate ways to analyze this effective field theory.
Firstly, to carry out the configuration averages over realizations of the random fields, we introduce $n$ replicas of each field. % (with replica indices indicated by $c$ and $c^\prime$).
%, with the understanding that we are to take the $n \to 0$ limit when computing physical properties.
The replicated field theory can then be used directly to generate the cumulant expansion,\cite{tarjus2006} or in the conventional manner, by taking  the $n \to 0$ limit when computing physical properties.
%%
% There can be subtle problems with results computed using the replica trick in spin-glasses but in the present case  that relates to random-field systems, replicas are only used to  generate configuration-averaged cumulants and there is no  spontaneous replica symmetry breaking to worry about.

Since the CDW never orders, it is also reasonable to treat the fluctuations of $ \psi$ in a self-consistent Gaussian approximation - this approximation becomes exact (at least in the loose sense commonly used in the field) in the limit $N\to\infty$.  The fluctuations of $\zeta$  do not involve any broken symmetries, and so to the same level of approximation, these can be treated in a saddle-point approximation, yielding the self-consistency equation in terms of $\mu\equiv 2i\zeta/\sqrt{N}$,
%\be
%\mu/2U +\Lambda=  T\big[A_1(\mu+{\cal N})+A_1(\mu-{\cal N})\big]+\sigma^2\big[A_2(\mu+{\cal N})+A_2(\mu-{\cal N})\big]
%\label{mu}
%\ee
which in the hard-spin limit ($U\to\infty$) becomes
\be
\Lambda=T\big[A_1(\mu+{\cal N})+A_1(\mu-{\cal N})\big]+\sigma^2\big[A_2(\mu+{\cal N})+A_2(\mu-{\cal N})\big].
\label{hardmu}
\ee
with $A_p$ given in Eq. (\ref{Am}). Notice that this constraint imposes a physically appropriate sum-rule on the integrated scattering intensity, $\int d\vec k S(\hat e_xQ+\vec k)+ \int d\vec k S(\hat e_yQ+\vec k)=(2\pi)^3 \Lambda $.

Although not necessary (see below), we can similarly evaluate the nematic order parameter approximately  directly from the saddle-point equation for $\phi$ in the limit $n\to 0$:
\be
\mathcal{N}/(2\Delta)=T\big[A_1(\mu-{\cal N})-A_1(\mu+{\cal N})\big]+\sigma^2\big[A_2(\mu-{\cal N})-A_2(\mu+{\cal N})\big]
\label{calN}
\ee
This relates the nematicity to the difference in the integrated scattering intensities, $\mathcal{N}(2\pi)^3=2\Delta\big[\int d\vec k S(\hat e_yQ+\vec k)- \int d\vec k S(\hat e_xQ+\vec k)\big]$.

Fig. 1 was obtained by numerically solving the  self-consistency equations for a lattice version of the same model, for the case $V_z = 0.01\kappa_{\parallel}, \kappa_{\perp} = 0.98\kappa_{\parallel}$.  For $V_z > 0.37 \kappa_\|$ and for $\sigma=0$, there is a single first-order transition from a stripe-ordered phase to the disordered phase with no intermediate nematic phase, but for non-zero $\sigma$ the stripe phase is  replaced by a nematic phase, although for weak enough disorder, the nematic transition is now first-order.  So long as $V_z \neq 0$, the solution obtained in this way is qualitatively reasonable;  however, while for non-zero $\sigma$, %the self-consistent solution  has
$\mu-|{\cal N}|>0$, % at all $T$,
which rightly implies that there is no stripe ordered phase,  we   obtain a solution with non-zero ${\cal N}$ for low enough $T$, even in the 2D limit $V_z\to 0$ where such a state is forbidden on general grounds. This is an artifact of the mean-field, saddle-point approximation for the nematic field.

In the Supplemental Material, we treat the effective field theory for $\phi$ more accurately.  Specifically, we show that upon integrating out the CDW fluctuations, the replicated field theory for $\phi$ is of  the same form as the replicated field theory of the RFIM.  At $T=0$ and in the limit of weak disorder, we can similarly map a correspondence between the two models by identifying the domain-wall energies and the mean-square disorder strength. These two exercises make explicit what is apparent by symmetry -- that the problem of nematic ordering is equivalent to the ordering of the RFIM.  The two qualitatively interesting aspects of this correspondence are that
\be
b^{eff} \sim |b_x|^2 - |b_y|^2; \ \ {\rm and} \ \sigma^{eff} \sim \sigma^2 \sqrt{A_4(\mu)},
\label{beff}
\ee
where $b^{eff}$ and $\sigma^{eff}$ are, respectively, the  uniform component and the root mean-squared variations of the effective magnetic fields which appears in the RFIM.  Importantly, this means that if the disorder is weak ($\sigma$ is small), the effective disorder felt by the nematic component of the order parameter is parametrically smaller still.  The mapping between the two models  permits one  to connect problems of vestigial nematic ordering to the rich, and well studied phenomenology of the RFIM.\cite{natterman}

{\bf Some implications for experiments in the cuprates:}
Clear evidence of the growth of short-range correlated CDW order in the pseudo-gap regime of the phase diagram has been found in a large number of experiments in multiple families of hole-doped cuprates.
\footnote{For reviews see \cite{rmp} and \cite{vojta2009};  for an update, see \cite{fradkin2012}.}
To interpret their significance, one would like to extrapolate the results to %infer something about the electronic structure of
an ``ideal hole-doped cuprate," one without quenched disorder and without structural peculiarities which lower the symmetry of the problem.  %It is by now clear that there is a
At least, the existing observations make clear that there is a
ubiquitous tendency to charge order with a well-defined period $\lambda$ which is a few times the lattice constant.  $\lambda$ depends on the doping concentration and certain structural details, presumably indicating that the electron-phonon coupling plays a role in  determining some aspects of the CDW order.  The preferred orientation of the CDW is always along the Cu-O bond ($x$ and $y$) directions. % (or, in the case of orthorhombic LSCO, almost along these directions).
\footnote{In LSCO, the density wave ordering vector is rotated slightly from the x and y directions, but this is an unimportant detail for present purposes, which is a necessary corollary\cite{robertson2006} of the particular orthorhombic structure of that material.}

However, there is no consensus about whether, in the absence of quenched randomness, the CDW order within each plane would be dominantly striped  ($\Delta >0$ ) or  checkerboard ($\Delta < 0$), whether the CDW order would be static (long-range ordered, $\Lambda>0$) or fluctuating (short-range correlated, $\Lambda<0$), and indeed whether the CDW phenomena seen in different cuprates are  siblings or distant cousins.
 As discussed previously in Refs. \cite{robertson2006,maestro2006} in the context of STM studies of the cuprates, in the presence of substantial disorder ($\sigma$ not  small) the structure factor itself   typically does not differ greatly between a ``failed'' stripe phase ({\it i.e.} with $\Lambda>0$ and $\Delta >0$) and a failed checkerboard phase ({\it i.e.} with $\Lambda>0$ and $\Delta <0$), nor whether the disorder is pinning what would otherwise be fluctuating order ($\Lambda<0$) or breaking up into domains what would otherwise be long-range CDW order ($\Lambda>0$).  To see this, consider the expression for the structure factor in Eq. (\ref{structure});  it has no explicit dependence on either $\Lambda$ or $\Delta$, but rather depends on them only implicitly through the self-consistency equations for $\mu$ and ${\cal N}$.  Because quenched disorder absolutely precludes  long-range CDW order, $\mu > |{\cal N}|$ independent of $\Lambda$;  only by approaching the limit of vanishing disorder would it be possible to distinguish unambiguously whether the correlation length, $1/\xi =\sqrt{\kappa(\mu-|{\cal N}|)}$ $(\kappa_{\perp} = \kappa_{\parallel} \equiv \kappa)$, is finite because of disorder or because of thermal or quantum fluctuations.  Moreover,  even in the presence of orthorhombicity ($|b^{eff}|>0$) or spontaneous nematic symmetry breaking (both of which result in ${\cal N}\neq 0$), so long as $1/\xi$ is not too long (or, more precisely, so long as $\mu \gg |{\cal N}|$), the structure factor %is still  presents an  has approximate
 only breaks $C_4$ symmetry weakly.

It has been suggested that stripe and checkerboard order can be distinguished by studying the structure factor at harmonics of the ordering vector;  for example, while either a stripe or a checkerboard ordered system with multiple macroscopic domains %of either orientation
 would exhibit equal peaks at $2Q\hat x$ and $2Q\hat y$,  only the checkerboard state would exhibit a second harmonic peak at $\vec Q_{xy}=Q(\hat x + \hat y)$.  This distinction does not
 %however,
 pertain to an uncondensed CDW:  Peaks in $S(\vec q)$ %occur  to the extent that non-linear terms in the effective action, proportional to higher than quadratic powers of $\psi_{\alpha}$, are taken into account. %In particular,
 at harmonics of the fundamental ordering vector
 arise as composites of the fundamental fields.
 %For instance, t
 The leading contribution %to the structure factor
 near the second-harmonic $\vec Q_{xy}=Q(\hat x + \hat y)$ is given by
\bea
S(\vec Q_{xy} + \vec k) &&\sim %g_{xy}
\int \frac {d\vec q}{(2\pi)^3} S(Q\hat x + \vec q)S(Q\hat y + \vec k-\vec q); %\nonumber \\ && \Longrightarrow \ *** \sigma^4 \ {\rm as}  \ {T\to 0}
\eea
%where in the final expression we have taken the low $T$ limit where thermal fluctuations can be neglected.
 %Although in a powered pattern of a CDW phase with long-range order, the presence of absence of this harmonic could be used to distinguish a stripe from a checkerboard phase, where long-range order has been suppressed by quenched randomness, even a measurement of the strength of this harmonic does
this does not  distinguish between the two cases any better than do the fundamentals.  %(For an explicit derivation, see Eq. \ref{**} in the Supplemental Material.)

 Thus, even though CDW formation is probably the fundamental ordering phenomenon involved, the nature of the ``ideal phase diagram" may be more directly inferred by studying the vestigial order.  If within the Cu-O planes, evidence can be adduced for the existence of  long-range nematic order, this likely implies that the ideal system would have long-range stripe order ($\Delta >0$ and $\Lambda > 0$), both because nematic order is a natural consequence of the existence of a striped ground state, and because even in the absence of quenched randomness, $T_{nem}$ is never much above $T_{str}$.  If, considering the effects of interplane interactions, the striped ground-state is gyrotropic\cite{hosur2013} (analogous to a cholesteric liquid crystal), for instance if the stripe orientation defines a handed spiral from plane to plane, then experiments which detect vestigial gyrotropic order can likewise be interpreted as evidence of a stripe ordered ground state in the ideal limit.

 Compelling evidence\footnote{For a contrary opinion, see \cite{ali}.} of nematic order within the Cu-O plane in BSCCO has been obtained from STM studies in Refs. \cite{howald2003,lawler2010,mesaros2011}.  In YBCO (in which the native orthorhombicity of the lattice complicates the analysis), evidence of nematic order has been inferred from transport anisotropy\cite{ando2002,daou2010}, from a strongly $T$ dependent growth in the anisotropy of the magnetic structure factor measured in neutron scattering\cite{hinkov2008} (although in a regime of doping below that in which X-ray evidence of CDW correlations has been found), and, for doping concentrations with the Ortho II structure, directly from anisotropy in the charge structure factor itself. \cite{mook2000,blackburn2012,keimer2013}
 %at one particular doping concentration (corresponding to the Ortho II structure)\cite{mook2000,blackburn2012,keimer2013}.
 In LBCO\cite{fujita2004,hucker2011}, because the LTT crystal structure produces a strong, explicit $C_4$ symmetry breaking in each plane, it is possible to unambiguously identify the charge-order as consisting of  stripes that rotate by $\pi/2$ from plane to plane\cite{hucker2011}.
% \footnote{Presumably, in YBCO in a high magnetic field, where NMR evidence of unidirectional long-range CDW order has been obtained, this reflects a combination of a commensurate lock-in of the CDW and a correspondingly strong effect of the crystalline orthorhombicity.}
   Moreover, in all three of these materials, as well as in Hg-1201, the onset of a spontaneous Kerr signal\cite{xia2008,he2011,karapetyan2012,karapetyan2013} below an onset temperature which correlates with the onset of other measures of the growth of local CDW correlations, has been tentatively identified\cite{hosur2013} as indicating the onset of gyrotropic order in all these materials.

 There are several sorts of experimental protocol that could  more unambiguously test for   nematic order.  As was done in the case of the Fe-based superconductors in Ref. \cite{kuo2013}, the ideal experiments would involve measuring an electronic property  which is odd under $C_4$ rotation (and hence proportional to ${\cal N}$) as a function of uniaxial strain to look for evidence of a divergent differential nematic susceptibility.  Particularly interesting would be NMR measurements of the quadrapolar broadening of the in-plane O lines, following along the lines undertaken (in orthorhombic YBCO in the absence of applied strain) in Ref. \cite{julien2013}  %There, the average quadrapolar shift, $\nu_a\propto\overline{\rho_a}$, and (by measuring the inhomogeneous broadening of the line) its mean-square variance $(\delta\nu_a)^2\propto \overline{ (\delta\rho_a)^2}$ were determined separately for
%the planar O's on x-directed and y-directed bonds, referred to  as O$_a$ with $a=$2 and 3, respectively.   ( %$ \overline{\rho_a}$ and $  \overline{ (\delta\rho_a)^2}$ are defined in
%See Eq. \ref{deltarho}.)  Consistent with our expectation based on Eq.\ref{deltarho}, the  $T$ dependent growth of $  \overline{ (\delta\rho_a)^2}$ was shown in Ref. \cite{julien2013} to correlate with other measures (from Kerr effect and x-rays) of the growth of local CDW order.  A $T$ dependent difference, $  \overline{ (\delta\rho_x)^2}-  \overline{ (\delta\rho_y)^2}$, was noted there as well, as possible evidence of nematicity, but because of the orthorhombic structure of YBCO, it was rightly concluded that this evidence was not conclusive;  measuring this same quantity as a function of uniaxial strain could remedy this problem.
(where a $T$ dependent difference $\overline{<|\delta \rho_x|^2 - |\delta \rho_y|^2>}$ was already noted in \cite{julien2013} as possible but inconclusive evidence of nematicity).

There are also numerous dynamical implications of the correspondence between the nematic order parameter and the RFIM.  Characteristic features of the expected hysteresis and noise implied by this correspondence\cite{carlson2006,phillabaum2012} have been reported in mesoscale samples of YBCO.\cite{bonetti2004}  Repeating this same sort of experiment, but with controlled manipulation of a uniaxial strain, will likely open up  other routes to study vestigial nematic order.

However, in light of the clear evidence of ubiquitous CDW correlations with significant short-range-order (largish $\xi$) and the growing evidence of nematic order, {\it it is reasonable to suppose that, absent quenched randomness, a state with long-range stripe-order, probably with a three dimensional structure that defines a handed spiral, would be found below a transition temperature, $T_{str}$, which is in the neighborhood of that at which X-rays see an onset  of short-range CDW order in the actual materials.}  There is also significant evidence supporting the notion that important fluctuation effects in the cuprates are associated with the existence of a quantum critical point (of a still to be agreed upon nature) under the superconducting dome;  a corollary of the above analysis is that, given that disorder is always relevant, {\it this quantum critical point cannot  be associated with the onset of a putative translation symmetry breaking transition, but could still be related to the onset of nematic order.}

\acknowledgements{  We acknowledge important conversations with Sri Raghu, Aharon Kapitulnik, John Tranquada, Ian Fisher, Michel Gingras, and Eduardo Fradkin.  }

\bibliography{DisorderCDW}

\end{document}

%\centerline{ \bf
\title{\bf
Supplemental Material for ``Quenched disorder and vestigial nematicity in the pseudo-gap regime of the cuprates''}
\author{Laimei Nie$^1$, Gilles Tarjus$^2$ and S. A. Kivelson$^1$}
\affiliation{1) Department of Physics, Stanford University, Stanford, California 94305, USA}
\affiliation{2) LPTMC, CNRS-UMR 7600, Universit\'e Pierre et Marie Curie, 4 Place Jussieu, 75252 Paris cedex 05, France}

\maketitle
%\section{Supplemental Material}
%subsection{Mapping to RFIM}
\section{Lattice Model}
To provide an explicit ultraviolet cutoff for integrals in a way that is guaranteed to respect the underlying lattice symmetries, we have defined a
 lattice version of Hamiltonian (Eq. (6) in the main text), which we have used when obtaining explicit numerical solutions of the model:
\bea
\label{Hlattice}
H &=& -J \sum\limits_{\langle \vec r, \vec r' \rangle} \sum_m \left[ \psi^{\dagger} (\vec r,m) \psi(\vec r',m) + \rm C.C. \right] \nonumber\\
  &-& J' \sum\limits_{\vec r,m}  \left[ \psi^{\dagger}(\vec r,m) \tau  \psi(\vec r+  \hat x,m) -  \psi^{\dagger}(\vec r,m) \tau  \psi(\vec r+ \hat y,m)
  + \rm C.C. \right] \nonumber \nonumber\\
  &+&  \frac{U}{2N}\sum\limits_{\vec r,m} \left[ \psi^\dagger(\vec r,m) \psi(\vec r,m) - N\Lambda \right]^2  -\frac{\Delta}{2N} \sum\limits_{\vec r,m}
  \left[ \psi^{\dagger}(\vec r,m) \tau  \psi(\vec r,m) \right]^2\nonumber\\
  &-& V_z\sum\limits_{\vec r, m} \left[\psi^{\dagger}(\vec r, m) \psi(\vec r,m+1) + \rm C.C. \right] \nonumber\\
  &-& \sum\limits_{\vec r,m} \left[ h^{\dagger}(\vec r,m)\psi(\vec r,m)+\rm C.C. \right]
\eea
where $\psi_{\alpha,j}$ is a two-index 2 $\times\ N$ component field, where $\alpha=$ x, y, refers to the direction of the CDW and for $N=2$, $j=1$, 2 refers
to the real and imaginary parts of the amplitude,
\be
J = \frac{1}{2}(\kappa_{\parallel}+\kappa_{\perp}), \quad J' = \frac{1}{2}(\kappa_{\parallel}-\kappa_{\perp}), \quad \tau =  \left( \begin{array}{cc}
\mathbb{I}_{N \times N} &   \\
 & \mathbb{-I}_{N \times N}
\end{array} \right).
\ee
The vector $\vec r$ denotes the position in a given layer ($x,y$ plane) and $m$ labels the layers ($z$ axis). When there is no ambiguity, we
will use the notation $\mathbf r \equiv (\vec r, m)$ in the following.
The $Z_2$ symmetry of the model under $x\to y$, $y\to -x$ and $\psi_{x,j}\to \psi_{y,j}$ and $\psi_{y,j}\to-\psi_{x,j}$ represents
the $C_4$ symmetry of the physical system, while the $SO(N)$ rotational symmetry represents a generalized  translational symmetry.
(In the physical $SO(2)$ case, the two components of $\psi_{\alpha,j}$ correspond to the real and imaginary parts of the complex
CDW amplitude, $\psi_\alpha$ defined in Eq. (1) of the main text.)

Henceforth, we will consider the model in the limit $U\to\infty$, where the term proportional to $U$ is omitted,  and instead  $\psi$ is
subjected to the hard-spin constraint, $\psi^\dagger\psi =N\Lambda$, which we enforce
by introducing the Lagrange-multiplyer field $\zeta(\mathbf r)$.  We also perform a Hubbard-Stratonovich
transformation which introduces the nematic field $\phi(\mathbf r)$ to replace the quartic term $\Delta$.  The
 Hamiltonian then reads
\bea
H[\psi,\phi,\mu,h] &=& -J \sum\limits_{\langle \vec r, \vec r' \rangle}\sum_m  \left[ \psi^{\dagger} (\vec r,m) \psi(\vec{r'},m)
+ \rm C.C. \right] \nonumber\\
  &-& J' \sum\limits_{\mathbf r}  \left[ \psi^{\dagger}(\mathbf r) \tau  \psi(\mathbf r+  \hat x) -  \psi^{\dagger}(\mathbf r) \tau  \psi(\mathbf r+ \hat y)
  + \rm C.C. \right] \nonumber \nonumber\\
  &+&  i\sum\limits_{\mathbf r}\zeta(\mathbf r) \left[ \psi^\dagger(\mathbf r) \psi(\mathbf r) - N\Lambda \right]
  +\frac 1 {\sqrt{N}}\sum\limits_{\mathbf r} \phi(\mathbf r)\left[\psi^{\dagger}(\mathbf r) \tau \psi(\mathbf r)  \right] +
  \frac{1}{2\Delta}\sum\limits_{\mathbf r} \phi^2(\mathbf r)\nonumber\\
  &-& V_z\sum\limits_{\vec r, m} \left[\psi^{\dagger}(\vec r, m) \psi(\vec r,m+1) + \rm C.C. \right] %\nonumber\\
 % &+&
 - \sum\limits_{\mathbf r} \left[ h^{\dagger}(\mathbf r)\psi(\mathbf r)+\rm C.C. \right]
\eea

\section{Replicas and the configuration average}

To better exhibit the statistical symmetries of the model, we introduce $n$ replicas of the system.  This allows us to define an effective,
translationally invariant  model in which the averages over the random fields have been explicitly performed:
\be
\exp\big(-\beta H_{rep}[\{\psi^{(a)},\phi^{(a)},\zeta^{(a)}\}]\big)\equiv\overline{\exp(-\sum_{a=1}^n  H[\psi^{(a)},\phi^{(a)},\zeta^{(a)};h] )}
\ee
where $h_{\alpha j}$ are Gaussian random variables with
\be
\overline{h_{\alpha j}(\mathbf r)}=0, \ \ \overline{h_{\alpha j}(\mathbf r)h_{\alpha^\prime j^\prime}(\mathbf{r'})} = \sigma^2 \delta_{\alpha,\alpha^\prime}
\delta_{j,j^\prime}\delta_{\vec r,\vec{r'}}\delta_{m,m'},
\ee
and hence
\be
 H_{rep}[\{\psi^{(a)},\phi^{(a)},\zeta^{(a)}\}] = \sum_a H[\psi^{(a)},\phi^{(a)},\zeta^{(a)};0] -\frac{\beta \sigma^2}{2} \sum_{a,a^\prime}
\sum_{\mathbf r} \big [ \psi^{(a)\dagger}(\mathbf r)   \psi^{(a^\prime)}(\mathbf r)+  % \psi_{y,a}(\mathbf R)\cdot  \psi_{y,b}(\mathbf R)
{\rm C.C.} \big].
\ee
%. As known in the context of the critical behavior of the RFIM,\cite{tarjus-tissier,tissier-tarjus}
To focus on the nematic order parameter itself, we formally define the effective Hamiltonian expressed in terms of the replica nematic fields
$\phi^{(a)}$ alone by integrating out the remaining fields,
\be
\label{eq_replicated_action_MSA}
\exp\left(-\beta H_{eff}[\{\phi^{(a)}\}]\right)\equiv %\exp\left[-\frac{\beta N}{16\Delta}\sum_{a} \sum_{\vec r}\phi^{(a)}(\vec r)^2\right]
   \int \prod_{a=1}^n d\zeta^{(a)}% e^{\beta N\Omega \sum_a \mu_a}
\mathcal D {\psi}^{(a)} \exp\big(-\beta H_{rep}[\{\psi^{(a)},\phi^{(a)},\zeta^{(a)}\}] \big).
\ee
Because of the Yukawa-like  coupling between  $\zeta$ and $\psi$, this formal process cannot be implemented exactly.  However, we can evaluate the
$\zeta$ integral in saddle-point approximation, which is exact in the large $N$ limit;  this is equivalent to replacing the hard-spin constraint by the
mean ``spherical'' constraint
\be
<\psi^{(a)\dagger}(\mathbf r)\psi^{(a)}(\mathbf r)>=\Lambda N,
\label{constraint}
\ee
which  serves as an implicit equation for the  saddle-point values of $\zeta^{(a)}(\mathbf r)=-i(\mu_a+2J+V_z)$, where $\mu_a$ is a constant in space.
Now, the integral over the CDW fields, $\psi^{(a)}$, is straightforward, since they are Gaussian and always massive,
\bea
\label{eq_replicated_action_MSA_trlog}
 H_{eff}[\{\phi^{(a)}\}]
&= &\frac 1  {2\Delta}\sum_{a} \sum_{\mathbf r}\phi^{(a)}(\mathbf r)^2
+\frac{TN}{2}\sum_{\alpha=\pm}
{\rm Tr}\big\{\ln\big(T\tilde\mathbf{\cal G}^{-1}[\{\phi^{(a)}\};\alpha]\big)\big\} \nonumber\\
  &-& N\Lambda \sum_a\sum_{\mathbf r}(\mu_a+2J+V_z)
\eea
where $\tilde\mathbf{ \cal G}^{-1}[\{\phi^{(a)}\};\pm]$ is a matrix in replica indices and position such that
\begin{equation}
\label{eq_operatorC}
\tilde{\cal G}^{-1}_{a \mathbf r,a^\prime \mathbf{r'}}[\{\phi^{(b)}\};\pm]
= \Big [ \tilde G^{-1}_{\mathbf r,\mathbf{r'}}(\mu_a;\pm)
\ \pm \frac {\phi^{(a)}(\mathbf r)} {\sqrt{N}} \delta_{\mathbf r,\mathbf{r'}}\Big ] \delta_{a,a^\prime}
-  \beta\sigma^2\
\delta_{\mathbf r,\mathbf{r'}}\ ,
\end{equation}
where we have used the notation $\{\phi^{(b)}\}$ to stress that $\tilde{\cal G}^{-1}_{a,a'}$ depends on all replicas fields (this is also true
for $\mu_a$ which depends on all $\{\phi^{(b)}\}$'s through the mean spherical condition). Furthermore,
\bea
\tilde G^{-1}_{\mathbf r,\mathbf{r'}}(\mu;\pm)=&&
- \frac{(J\pm J^\prime)}2 [\delta_{\mathbf r-\mathbf{r'} ,\hat x}+\delta_{\mathbf r-\mathbf{r'} ,-\hat x}] +
-\frac{(J\mp J^\prime)}2 [\delta_{\mathbf r-\mathbf{r'} ,\hat y}+\delta_{\mathbf r-\mathbf{r^\prime} ,-\hat y}]  \nonumber \\
&& - \frac {V_z} 2[\delta_{\mathbf r-\mathbf{r^\prime} ,\hat z}+\delta_{\mathbf r-\mathbf{r^\prime} ,-\hat z}]
+(\mu+2J+V_z) \delta_{\mathbf r,\mathbf{r^\prime}}\ .
\label{tildeG}
\eea
Exploiting the translational symmetry of the replicated model, we can obtain the Fourier transform of $\tilde G$:
\be
G(\mathbf k;\mu;\pm)^{-1} = 2{(J\pm J^\prime)}\sin^2(k_x/2)
+2(J\mp J^\prime)\sin^2(k_x/2) +2V_z\sin^2(k_z)
+\mu\ ,
\label{G}
\ee
where $G$ is the lattice version of the corresponding quantity defined in Eq. (3) of the article.

The formal expression for $H_{eff}$ is generally extremely complicated. It can be expanded in increasing number of sums
over replicas to generate a cumulant expansion\cite{tarjus2008} and can further be
expanded in gradients of the fields $\phi^{(a)}$, assuming that the latter are slowly varying in space.  In the
case where we completely neglect the spatial variation of $\phi^{(a)}$, we can define ${\cal N}_a\equiv \phi^{(a)}/\sqrt{N}$, and
$\tilde {\cal G}$ can be diagonalized by Fourier transform, yielding
\bea
{\cal G}_{aa^\prime}(\mathbf k;\{{\cal N}_b\};\pm)= G(\mathbf k;\mu_a\pm {\cal N}_a;\pm)\delta_{a,a^\prime}
+ \beta \sigma^2 \frac{G(\mathbf k;\mu_a\pm {\cal N}_a;\pm) G(\mathbf k;\mu_{a^\prime}\pm {\cal N}_{a^\prime};\pm)}
{1-\beta \sigma^2\sum_b G(\mathbf k;\mu_b\pm {\cal N}_b;\pm)} \,.
\label{calG}
\eea
Under these circumstances,
\bea
&&H_{eff}[\{\sqrt{N}{\cal N}_a\}] =\nonumber \\&& N V \left\{\sum_a \left [\frac {{\cal N}_a^2}{2\Delta} - \Lambda (\mu_a+2J+V_z)\right ]
- \frac T 2 \sum_{\alpha=\pm}\int \frac { d^{3} k}{(2\pi)^3} {\rm Tr}
\big \{ \ln\big(T{\cal G}[\mathbf k;\{\mathcal N_a\};\alpha]  \big)\big\}\right\}
\eea
where $V=\sum_{\vec r} 1$ is the volume and the trace, now, is only over the replica index.  After expanding in increasing number of sums over
replicas, we obtain
\bea
&&\frac{H_{eff}[\{\sqrt{N}{\cal N}_a\}]}{N V}=\nonumber \\&& \sum_a \left \{\frac {{\cal N}_a^2}{2\Delta} - \Lambda \big (\mu[{\cal N}_a]+2J+V_z \big )
- \frac T 2 \sum_{\alpha=\pm}\int \frac { d^{3} k}{(2\pi)^3}\Big [\ln\big(T G(\mathbf k;{\cal N}_a;\alpha)\big )
+\beta \sigma^2G(\mathbf k;{\cal N}_a;\alpha)\Big ]\right \}\nonumber \\&&
-\frac{\beta \sigma^4}{4} \sum_{a,a'} \sum_{\alpha=\pm}\int \frac { d^{3} k}{(2\pi)^3} G(\mathbf k;{\cal N}_a;\alpha)G(\mathbf k;{\cal N}_{a'};\alpha)
+\mathcal O(\sum_{a,a',a''})
\label{eq_expansion_replicas}
\eea
where we have defined for convenience $G(\mathbf k;{\cal N}_a;\pm)\equiv G(\mathbf k;\mu[{\cal N}_a]\pm {\cal N}_a;\pm)$ and $\mu[{\cal N}_a]$
is solution of the saddle-point equation at the lowest order in the number of sums over replicas:
\be
\Lambda = T \sum_{\alpha=\pm}\int \frac { d^{3} k}{(2\pi)^3} \left [G(\mathbf k;{\cal N}_a;\alpha)
+  \sigma^2 G(\mathbf k; {\cal N}_a;\alpha)^2 \right ]\,.
\label{eq_saddlepoint_0}
\ee
Note that when all replica nematic fields are equal, ${\cal N}_a=\mathcal N$, the above expansion in Eq. (\ref{eq_expansion_replicas}) is
equivalent to an expansion in powers of the number of replicas $n$ and one recovers the standard replica trick when $n\rightarrow 0$.

The replicated theory makes manifest the statistical symmetries of the problem.  Clearly, $H_{eff}$ in Eq. (\ref{eq_replicated_action_MSA_trlog})
is translationally invariant.  However, the index $\pm$ in $\tilde G$ brings on an explicit dependence on spatial orientation;  for $+$ the preferred
axis in the $x$ direction and for $-$ it is in the $y$ direction.  Thus, $H_{eff}$ has a sort of ``spin-orbit coupling,'' such that it is not invariant under
$C_4$ spatial rotation nor any transformation of the order parameter alone.  Moreover, because of the coupling between different replicas generated
by the $\sigma$ dependent terms, no transformation that acts on a subset of replicas leaves $H_{eff}$ invariant; this is the property that identifies the
problem as a random-field problem.  $H_{eff}$ is invariant under  the discrete rotation  $\phi^{(a)}(\vec r,m) \to -\phi^{(a)}(\vec{r^\prime},m)$ with
$x^\prime=y$ and $y^\prime=-x$.  This is the symmetry that  identifies the problem as a version of the Ising model.  (The
model is also invariant under the mirror-plane transformation $\phi^{(a)}(\vec r,m) \to -\phi^{(a)}(\vec{r^\prime},m)$
with $x^\prime=y$ and $y^\prime=x$.)

\section{Relation to the RFIM}

To establish the relation between $H_{eff}$ and the RFIM, we perform the same sort of analysis for the RFIM.  We start with  a general Ising
ferromagnet in a random field,
\be
\beta H_{RFIM}[S]=-\frac 12 \sum_{ij}S_iK_{ij}S_j -\beta\sum_i H_iS_i
\ee
where $S_i=\pm 1$, $K_{ij}\geq 0$, and $H_j$ a Gaussian random variable with zero mean.  This can be recast in terms of real
scalar fields $\Phi_i$ by a series of transformations discussed in Ref. \cite{amit} as
\bea
\beta \tilde H_{RFIM}[\Phi]=&&\frac 12\sum_{ij}\Phi_iK_{ij}\Phi_j -\sum_i\ln\big[\cosh(\sum_j 2K_{ij}\Phi_j)\big] \nonumber \\
&&-\beta \sum_i H_i\Phi_i +\frac {\beta^2}2
\sum_{ij} %\frac {H_i H_j} {K_{ij}}.
H_i K_{ij}^{-1}H_j
\label{rfim1}
\eea
Here, the first two terms represent the effective Hamiltonian of the pure Ising ferromagnet, and the final term can be viewed
as a correction to the random field distribution.  Just as we did for the CDW model, we introduce $n$ replicas of the Ising fields, and
then perform the average over the random variables, resulting in
\bea
&&\beta H_{RFIM}^{eff}[\{\Phi^{(a)}\}]=\nonumber \\&& \sum_a\left\{\sum_{ij}\frac 12 \Phi_i^{(a)}K_{ij}\Phi_j^{(a)}
-\sum_i\ln\big[\cosh(\sum_j 2K_{ij}\Phi_j^{(a)})\big]\right\} -\frac{\beta^2}{2}
\sum_{a,a^\prime}\sum_{ij}\Phi_i^{(a)}D_{ij}\Phi_j^{(a^\prime)}
\label{rfim}
\eea
where
\be
D_{ij} = \overline{H_iH_j}
\ee
with the average performed over an ensemble that includes the effect of the final term in Eq. (\ref{rfim1}).

The symmetries of this problem are manifestly similar to those of $H_{eff}$.  Again, there is no symmetry under transformations which involve a
subset of the replicas.  Indeed, $H_{RFIM}^{eff}$ is invariant under all the same transformations as $H_{eff}$, but because the RFIM as defined
has no spin-orbit coupling, it has an additional invariance with respect to pure spatial transformations
of the type $\Phi^{(a)}(\mathbf r)\to \Phi^{(a)}(\mathbf{r^\prime})$.

An explicit correspondence between the two models can be made in different fashions in different parameter regimes
(compare for instance Eqs. (\ref{eq_expansion_replicas}) and (\ref{rfim}) when the field $\Phi$ is uniform).  For $T$ near to the nematic
ordering temperature, the effective Hamiltonian can be expanded in powers of the order parameter fields and their spatial
derivatives, and can be compared term by term.  To illustrate the point, we consider the terms in $H_{eff}$ to zeroth order in spatial
derivatives (i.e. evaluated for constant values of $\phi^{(a)}=\sqrt N \, \mathcal N_a$).
From Eq. (\ref{eq_expansion_replicas}) one easily derives
\bea
\beta H_{eff} = &&\sum_{\mathbf r}\left\{ \sum_a \Big [\frac {B_1}2 \mathcal N_a^2 + \frac{C_1}{4!}\mathcal N_a^4 \Big ]\right .
 -\frac 12 \sum_{a,a^\prime}\Big [  B_2 \mathcal N_a\mathcal N_{a^\prime} +C_2 \mathcal N_a^2\mathcal N_{a^\prime}^2\Big] \\ \nonumber
&&\left . + \frac {1}{3!} \sum_{a,a^\prime,a^{\prime\prime} }C_3 \mathcal N_a\mathcal N_{a^\prime}\mathcal N_{a^{\prime\prime}}^2
%+ \frac {1}{4!}\sum_{a,a^\prime,a^{\prime\prime},a^{\prime\prime\prime}  }C_4\mathcal N_a\mathcal N_{a'}\phi^{(a^{\prime\prime)}}
 %\phi^{(a^{\prime\prime\prime)}}
 \right\}
 +\ldots
\eea
where $\ldots$ indicates higher powers of the field and their derivatives and
\bea
&&\frac{B_1}{N} = \frac {\beta}{\Delta} - \mu''_0\Big (\beta \Lambda -\frac 12\sum_{\alpha=\pm}\int \frac { d^{3} k}{(2\pi)^3}
\big [G(\mathbf k;\mu_0;\alpha)+\sigma^2G(\mathbf k;\mu_0;\alpha)^2\big ]\Big )
\nonumber \\&&-\sum_{\alpha=\pm}\int \frac { d^{3} k}{(2\pi)^3}
\big [G(\mathbf k;\mu_0;\alpha)^2+\sigma^2G(\mathbf k;\mu_0;\alpha)^3\big ], \nonumber \\
&&\frac {B_2}{N}= \frac{\beta^2 \sigma^4}{2}  \sum_{\alpha=\pm}\int \frac { d^{3} k}{(2\pi)^3}
G(\mathbf k;\mu_0;\alpha)^4 \nonumber \\
&&\frac{C_2}{N}= 2 \beta^2 \sigma^4  \sum_{\alpha=\pm}\int \frac { d^{3} k}{(2\pi)^3}
G(\mathbf k;\mu_0;\alpha)^6
\eea
where $\mu_0$ is the solution of Eq. (\ref{eq_saddlepoint_0}) when $\mathcal N=0$ and
$\mu''_0=\partial^2\mu/\partial \mathcal N^2\vert_{\mathcal N=0}$. Moreover, $C_3\neq 0$ when $\sigma^2>0$

The corresponding expression for $H_{RFIM}^{eff}$ is of the same form, but with parameters
\bea
&&B_1^\prime =K\left(1-2 K \right)=\beta T_{MF}(1-\beta T_{MF}), \nonumber \\
&&B_2^\prime= \beta^2D\nonumber \\
&&C_2^\prime= C_3^\prime= \cdots = 0
\eea
where
\be
K\equiv \sum_jK_{ij}\equiv \beta T_{MF}/2, \ \ {\rm and} \ \ D\equiv \sum_j D_{ij}.
\ee
The expression of the other terms can be similarly obtained but are not particularly illuminating and are not given here.

There are some manifest, but ultimately unimportant differences in the structure of the two models.  Firstly, $C_p=0$ for all $p>1$ in the
standard RFIM. This is an artifact of the simple version of the model assumed;  random bond disorder (randomness in the values of $K_{ij}$)
would immediately generate a non-zero $C_2$ and a non-Gaussian distribution of the random fields as well as a combination of both random
bonds and random fields result in non-zero values for the other coefficients.  These terms are irrelevant for the universal physics at large scale.
A more subtle issue is that
$B'_1$ is independent of the disorder in the RFIM, while its counterpart depends implicitly on $\sigma$ for the CDW system;  again, this is a
peculiarity of the simple version of the RFIM considered, and the generic behavior (exhibited by the CDW model) would be generated by
an imperfectly Gaussian distribution of random fields.   While $B_1$, and $B_1^\prime$ both change sign at a non-zero mean-field transition
temperature, $T_{MF}$,  the $T$  dependence of $B_1^\prime$ is much
more complex than that of $B_1$; to make a precise correspondence between the models, the coupling constants entering
the RFIM would have to be $T$ and $\sigma$ dependent.

It is also possible to directly compare the two effective models in the limit $T\to 0$, with results analogous to those given above,
but we do not expand on this aspect here.

Despite the complexity that  accompanies any attempt to establish a precise mapping between the two
models, it is clear that the structure of the two models is sufficiently similar that one can adopt known results for the
RFIM qualitatively and even semi-quantitatively for the CDW system.

In the following sections we will treat the nematic order parameter in the saddle-point approximation as a way to illustrate our
conclusions by concrete results.  This is entirely analogous to treating the effective field theory for the RFIM at the same
level of approximation and could be replaced by more sophisticated treatments.

For the most part, the saddle-point solutions produce results that are qualitatively correct.  Of course, (as we shall see) it
produces mean-field exponents for various
critical properties, where non-trivial exponents would be expected in a more accurate treatment.  Moreover, nowhere does the
mean-field theory addresses the physics
of rare events (``droplets'') that lead to the extreme dynamical slowing down which is characteristic of the RFIM.

However, the most important failure of the mean-field treatment occurs in the in $d=2$ limit, $V_z=0$, where there is a
particular subtlety associated with the formation
of Imry-Ma domains - whereas the saddle-point equations admit a nematic phase at weak enough disorder in 2D, the
correspondence with the RFIM implies that instead
there should always be a finite nematic correlation length which in the weak disorder limit is exponentially long,
\be
\ln[\xi_{2D}] \sim  (\kappa/\sigma^{eff})^2
\ee
where $\xi_{2D}$ is the correlation length of the 2D RFIM with a random field  of RMS magnitude $\sigma^{eff}\sim \sigma^2/J$.
This subtlety, however, is less alarming than it seems at first, as it is eliminated by even extremely weak  3D couplings.  To make
an estimate of the way in which non-zero $V_z$ eliminates this 2D peculiarity, we estimate a length scale associated with small non-zero
$V_z$ in the following manner:  consider a block of $L\times L$ spins in a given plane and treat them as a single, block spin.  The
effective coupling between block spins within a plane is $JL$, while the effective coupling between planes is $V_zL^2$, so
for blocks of size $L=J/V_z$, the couplings become effectively isotropic, and 2D physics is no longer pertinent.
Thus, the physics of 2D Imry-Ma domains is negligible so long as $V_z > J/\xi_{2D}$.

\section{Mean-field solution}

\subsection{Saddle-point equations}

We now turn to the saddle-point, or mean-field solution of the problem.  For each replica, $\mu^{(a)}$ is determined
by the mean-spherical constraint,  Eq. (\ref{constraint}),
\be
\Lambda = \frac T V\sum_{\mathbf r}\bigg\{ \tilde {\cal G}_{\mathbf r a,\mathbf r a}[\{\phi^{(b)}\};-]
+\tilde {\cal G}_{\mathbf r a,\mathbf r a}[\{\phi^{(b)}\};+]\bigg\}
\label{saddle1}
\ee
while the saddle-point equations for the replicated field theory are given by
\be
\phi^{(a)}(\mathbf r) =  {T\Delta}{\sqrt{N}}\bigg\{ \tilde {\cal G}_{\mathbf r a,\mathbf r a}[\{\phi^{(b)}\};-] -
\tilde {\cal G}_{\mathbf r a,\mathbf r a}[\{\phi^{(b)}\};+]\bigg\}
\label{saddle2}
\ee
with $\tilde {\cal G}$ obtained from Eq. (\ref{eq_operatorC}). Note that the symmetry preserving state, $\phi^{(a)} =0$, is always a
solution of the set of equations,  (\ref{saddle1}) and (\ref{saddle2}).

There is no proof that the non-trivial solutions of these equations with lowest free energy are always homogeneous, but we will restrict
ourselves to this case.  Then, as before, defining
${\cal N}_a\equiv \phi^{(a)}/\sqrt{N}$, we can cancel the $N$ dependence of these equations
(making the $N\to \infty$ limit trivial to obtain, if we so desire). We are interested by the solution at the lowest order in the number of sums over
replicas, or equivalently by the limit $n\rightarrow 0$ (see above). In this case the saddle-point equations become
\bea
\label{saddle3}
\Lambda %=  && T \big\{  {\cal A}_{a,a}(-) +{\cal A}_{a,a}(+) \big\} \\
=T \big[  A_1(\mu_a-{\cal N}_a) +A_1(\mu_a+{\cal N}_a)\big]  +\sigma^2 \big[A_2(\mu_a-{\cal N}_a) +A_2(\mu_a+{\cal N}_a) \big]\big]
\nonumber
\eea
\bea
\label{saddle}
\frac{{\cal N}_a}{\Delta}%=  &&{T}\big\{ {\cal A}_{a,a}(-) -{\cal A}_{a,a}(+)\big\}\\
=T \big[  A_1(\mu_a-{\cal N}_a) -A_1(\mu_a+{\cal N}_a)\big]  +\sigma^2 \big[A_2(\mu_a-{\cal N}_a) -A_2(\mu_a+{\cal N}_a)\big]
\nonumber
\eea
where $\mu_a \equiv \mu[\mathcal N_a]$ (see also Eq. (\ref{eq_saddlepoint_0})) and
%${\cal G}$ and $G$ are  given in Eqs. (\ref{calG}) and (\ref{G}), respectively, and
%\be
%{\cal A}_{aa^\prime}(\pm) =\int \frac {d^{3}k} {(2\pi)^3}{\cal G}_{aa^\prime}(\mathbf k;\pm)
%\ee
\be
\label{Ap}
A_p(\mu) = \int \frac {d^{3}k} {(2\pi)^3}G(\mathbf k;\mu;\pm)G(\mathbf k;\mu;\pm)^{p-1}
\ee
with $G$ given in Eq. (\ref{G}). These equations are of precisely the form as the saddle-point equations given in Eqs. (9) and (10)
of the main text, with the the lattice propagator $G$ instead of the
continuum propagator in the definition of $A_p$.  The latter difference is convenient for numerical studies,
as no artificial cutoff needs to be introduced to perform the integrals (which are carried out for $\mathbf k$ in the first Brillouin zone).
Note that because of the integral over all $\mathbf k$, $A_p$ does not depend on the index $\pm$.
There is a separate, identical saddle-point equation for each value of the replica index, $a$, and, as is known from the mean-field
solution of the RFIM, no exotic spontaneous replica symmetry breaking is to be expected in this case. In the remainder of this section,
we explore the solutions of these saddle-point equations.

\subsection{Mean-field phase diagram}

%****ADDED BY SAK
The mean-field phase diagrams shown in Fig. 1 in the main text and below, are obtained by solving the saddle-loin  (mean-field) equations numerically in the $n\to 0$ limit (or, equivalently, in the replica symmetric case).  The most general form of these equation is
\bea
\label{meanfield}
\Lambda&&=|\Gamma|^2 +  T \big[  A_1(\mu-{\cal N}) +A_1(\mu+{\cal N})\big]  +\sigma^2 \big[A_2(\mu-{\cal N}) +A_2(\mu+{\cal N}) \big]
\\
{\cal N}/\Delta&&=  |\Gamma|^2+
T \big[  A_1(\mu-{\cal N}) -A_1(\mu+{\cal N})\big]  +\sigma^2 \big[A_2(\mu-{\cal N}) -A_2(\mu+{\cal N})\big] + b^{eff}
\nonumber
\eea
where $\Gamma=\overline{\langle \psi_x\rangle}$ is the magnitude of the CDW condensate (where we are still assuming that $\overline{\langle \psi_y\rangle}=0$), and $b^{eff}$ is a possible external symmetry breaking field (orthorhombicity) which (when positive) favors the nematic principle axis in the $x$ direction (positive ${\cal N}$).  Unless otherwise stated, we will always assume that the crystal has tetragonal symmetry, so $b^{eff}=0$ and nematicity arises solely as a consequence of spontaneous symmetry breaking.

\subsubsection{Clean limit $\sigma=0$}
In all the discussion in the main text, we have always assumed $\Gamma=0$, as it must be for $\sigma>0$ in $d \leq 4$.
To confirm this, note that the  spectrum of excitations about the saddle-point is given by Eq. (\ref{G}).
Because any phase with $\Gamma\neq 0$  breaks a continuous symmetry it must have a Goldstone mode;  thus, any phase with a non-zero value of $\Gamma$ must
have $\mu- |{\cal N}|=0$.  However, for
$\sigma^2$, this results in a divergent value of $A_2$ (in $d \leq 4$), and hence a violation of the hard-spin constraint.
This reflects the absence of continuous symmetry breaking in the presence of quenched randomness.

\begin{figure}
\centering
\includegraphics[width=10cm]{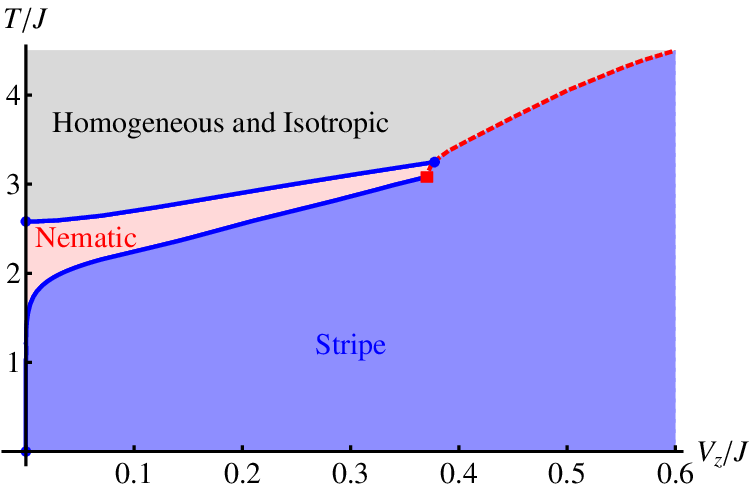}
\caption{The phase diagram in the clean limit ($\sigma=0$) as a function of $T$ and interplane coupling, $V_z$, with $J=1, J'=0.01,$ and $\Delta=0.25$ obtained by numerically solving Eqs. (\ref{meanfield}).  The solid and dashed lines represent, respectively, continuous and first-order phase transitions, the square a classical tricritical point, and the solid circle a critical end point. The phase boundary of the stripe phase has been shifted, for graphical clarity, since the nematic phase typically is confined to a still narrower range of $T$ than shown.}
\label{cleanphases}
\end{figure}

However, when we compute the phase diagram  in the clean limit shown in Fig. \ref{cleanphases}, we must include a non-vanishing $\Gamma$ at all temperatures below $T_{str}$. The continuous phase transitions in this diagram are straightforward to obtain directly from the self-consistency equations;  however, there are generally two distinct solutions to these equations in the vicinity of the first-order portions of the phase boundaries.  Thus, to determine the location of these boundaries, it is necessary to compute the Feynman variational free energy corresponding to each solution, and then favor the one with the lower free energy.  For small enough $V_z$ (i.e. for $V_z< 0.38 J$ in the case we have studied numerically, with $J'=0.01J$ and $\Delta=0.25J$), all the transitions are continuous, but for $V_z$ larger than a critical value at which there is a tricritical point, the stripe transition becomes first order.

\subsubsection{Phase diagram with disorder}

A non-zero nematic order parameter is possible, even with quenched randomness, for $d > 2$.   Indeed, it is
straightforward to see from Eqs. (\ref{saddle3}) and (\ref{saddle}) that at any temperature for which there is a
non-zero value of the nematic order parameter  at $\sigma=0$,  there
will still be a non-zero solution for small enough non-zero $\sigma$.  The proof of this assertion is particularly simple
for $T_{nem} >T> T_{str}$, where $A_p(z)$ are analytic functions in the neighborhood of $z=\mu(T,\sigma=0)\pm {\cal N}(T,\sigma=0)$.
It thus follows trivially that  both the
nematic order parameter and the ``mass'' of the CDW fluctuations (which determines the longest CDW correlation length
as shown in Eq. (\ref{xi}), below) are analytic functions of disorder strength:
\bea
\label{smallsigma}
&& \mu_-(T,\sigma)\equiv \mu(T,\sigma) - |{\cal N}(T,\sigma)|  =  \mu_-(T,0)  + {\cal O}(\beta\sigma^2) \\
&&|{\cal N}(T,\sigma)| = |{\cal N}(T,0)| - {\cal O}(\beta\sigma^2) \ \ {\rm for } \ \ T_{str} < T < T_{nem}\ . \\
\nonumber
\eea
Note that ${\cal N}$ is, by definition, the nematic order parameter and $\mu_-$ is a measure of how far the
system is from a CDW ordered state - below, we relate it to the CDW correlation length.

For $T< T_{str}$, the analysis is a bit more subtle, since $\mu_-(T,\sigma)  \to 0$ as $\sigma\to 0$.  The results,
moreover, depend on the asymptotic forms of
$A_p(\mu)$ at small $\mu$.  In $d=3$ the leading order behavior as $z\to 0$ is readily derived from the asymptotic expressions:
\bea
&& A_1(z)\sim A_1(0) - [A/(1-\alpha)]z^{1-\alpha} + \ldots \ \ {\rm and} \\
&&A_2(z) \sim Az^{-\alpha} + \ldots \  \ \ {\rm with}  \nonumber \\
&& \alpha = (4-d)/2 = 1/2 \ \ \ \ {\rm and} \ \ \ A^{-1}= {{2\pi}\sqrt{[J^2-(J^\prime)^2]V_z}},  \nonumber
\eea
from which it follows that Eq. (\ref{smallsigma}) is still satisfied, but with
\be
\mu_-(T,\sigma)  =  (1-\alpha)\beta\sigma^2+ {\cal O}(\beta^3\sigma^4) \ \ {\rm for} \ \ \beta\sigma^2 \ll T < T_{str}.
\ee
%(Note that, strictly speaking, the model we are considering has $d=3$, and therefore $\alpha=1/2$
%properly enters the true asymptotic analysis $\sigma$ analysis.
(Surprisingly, in the range of $T$ and $\sigma$ to which this applies, $\mu_-$ is a decreasing function of $T$ - since a
smaller $\mu_-$ implies a larger correlation
length, this corresponds to a range of temperatures in which the correlation length decreases with decreasing $T$!)
Manifestly, for fixed small $\sigma$, this expansion breaks down at low $T$, but  similar asymptotic analysis can be applied in
the limit of low $T$ and small $\sigma$
to obtain
\bea
&&\mu_-(T,\sigma)  \sim (A\sigma^2/\Lambda)^{1/\alpha} +\ldots \ \ {\rm for} \ \ T \ll \sigma  \ll T_{str}.\\
&&{\cal N} \sim \Lambda\Delta  -2TA_1(2\Lambda \Delta) - 2\sigma^2 A_2(2\Lambda \Delta) + \ldots
\eea
where $\ldots$ signifies higher order terms in both $T$ and $\sigma$.

 In a highly anisotropic system ($d\approx 2$), with $V_z\ll J$, there is an intermediate asymptotic regime in which
 $J \gg \beta \sigma^2 \gg V_z \gg J/\xi_{2D}$, in
 which the asymptotic forms of $A_p$  can be computed with $V_z=0$, in which case
 \bea
&& A_1(z)\sim A \ln[J/z] + \ldots \ \ {\rm and} \\
&&A_2(z) \sim A\mu^{-1} + \ldots \  \ \ {\rm with}  \nonumber \\
&& A^{-1}= {{4\pi}\sqrt{[J^2-(J^\prime)^2]}}.  \nonumber
\eea
In this limit, as well, Eq. (\ref{smallsigma}) governs the evolution  at small $\sigma$.

All together, independent of regime, the above analysis confirms, as shown in Fig. 1 in the text, that the nematic order parameter  is a
continuous function of disorder,
regardless of whether or not there is CDW order in the $\sigma\to 0$ limit.

Similar asymptotic analysis can be applied to determining the shape of the phase diagram.
%For non-zero $\sigma$, we find that the nematic transition is always continuous.
For small enough $V_z$, the nematic transition is continuous, so  we can identify $T_{nem}$ by equating the derivate with respect to
${\cal N}$ of the left and right sides
of Eq. (\ref{saddle}).  The critical value  $\mu_{c}\equiv\mu(T_{nem})$ is obtained as the solution of the implicit equation
\be
\Lambda\Delta A_2(\mu_{c})-A_1(\mu_{c}) =2\Delta\sigma^2\big[A_2^2(\mu_{c})-2A_1(\mu_{c})A_3(\mu_{c})\big]
\ee
in terms of which
\be
T_{nem}=\frac  {\Lambda-2\sigma^2A_2(\mu_{c})} {2A_1(\mu_{c})}.
\ee
%(In the case $V_z > \mu$, interplane coupling cuts off the divergence of $A_1$ but $A_2$ still diverges as $A_2 \sim |\mu|^{-1/2}$;
%this correctly reflects the fact
that breaking of $SO(N)$ symmetry is allowed in d=3 in the absence of disorder ($\sigma=0$), but not in the presence of disorder.)
Because $\mu_c$ is non-zero, all the dependence of the saddle-point equations on ${\cal N}$ is analytic for small ${\cal N}$.
Consequently, as in any other mean-field theory,
\be
{\cal N}(T) \sim {\cal N}_0\sqrt{[T_{nem}-T]/T_{nem}}
\ee
for $T_{nem} \gg [T_{nem} - T] > 0$.
$T_{nem}$ is a monotone decreasing function of $\sigma$ such that
\be
T_{nem} \to \Lambda/2A_1(\mu_c) \ \ {\rm as}\ \ \sigma\to 0
\ee
where $\mu_c$ is the solution of the implicit equation
\be
\Delta\Lambda=A_1(\mu_c)/A_2(\mu_c)
\ee
and
\be
T_{nem} \to 0 \ \ {\rm as}\ \ \sigma^2\to\sigma_c^2 = \Lambda /[2A_2(\mu_0)]
\ee
where $\mu_0$ is the solution of the implicit equation
\be
2\Delta \Lambda  = A_2(\mu_0)/A_3(\mu_0).
\ee
Note that these equations have a non-zero solution for any non-zero $\Delta$.

The phase diagram in Fig. 1 of the main text interpolates between these various asymptotic expressions, and was obtained by solving the self-consistency equations numerically.  Since we have focussed on relatively small values of $V_z$, all the transitions are continuous.  For larger values of $V_z$, where in the clean limit there is a single first order transition to a stripe ordered phase ({\it i.e.} for $V_z$ larger than the value at the critical end-point in Fig.\ref{cleanphases}, the nematic transition transition in the weak disorder limit is also   first-order.  We have not analyzed this limit extensively.

\subsection{The CDW structure factor}

The self-consistent fields, $\mu$ and ${\cal N}$, are the key quantities that determine the behavior of the response functions of the
system, as well as its thermodynamic state.
The  CDW structure factor, $S(\mathbf k)$ for $\mathbf k$ near the clean-limit ordering vectors, $Q\hat x$ and $Q\hat y$, is
expressed in terms of the propagator, $G(\mathbf k;\mu\pm {\cal N};\pm)$, in Eq. (\ref{G}).   The expected line shape consists of a sum of a
Lorentzian and a squared Lorentzian.  As a function of decreasing
temperature, the relative weight of the two factors shifts from being dominated by the former at high $T$ to being dominated
by the latter at low $T$.  From the
width of the peaks, one can extract a set of CDW correlation lengths (expressed in units of the lattice constant, as is appropriate for the lattice model in
Eq. (\ref{Hlattice}) - in terms of the original CDW, this lattice constant is a somewhat ill-defined ultra-violet cutoff which should be interpreted to be
something like the CDW wave-length.)  In general, there is an in-plane longitudinal and transverse correlation length, $\xi_L$ and $\xi_T$, as well as a
correlation length in the $z$ direction, $\xi_z$;  in a nematic state, all these
correlation lengths are different near the two ordering vectors.  Specifically,
\bea
\label{xi}
&&\xi_L(Q\hat x) = \sqrt{\frac {(J+J^\prime)}{2(\mu-{\cal N})}}, \ \  \xi_T(Q\hat x) = \sqrt{\frac {(J-J^\prime)}{2(\mu-{\cal N})}}, \ \ \xi_z(Q\hat x)
= \sqrt{\frac {V_z}{2(\mu-{\cal N})}}, \\
&&\xi_L(Q\hat y) = \sqrt{\frac {(J-J^\prime)}{2(\mu+{\cal N})}}, \ \  \xi_T(Q\hat y) = \sqrt{\frac {(J+J^\prime)}{2(\mu+{\cal N})}}, \ \ \xi_z(Q\hat y)
= \sqrt{\frac {V_z}{2(\mu+{\cal N})}}. \nonumber
\eea
The maximum scattering intensity is even more directly related to the self-consistent fields,
\be
S(Q\hat x) =\frac T {(\mu-{\cal N})}+ \frac {\sigma^2}{(\mu-{\cal N})^2}, \ \ S(Q\hat y) =\frac T {(\mu+{\cal N})}+ \frac {\sigma^2}{(\mu+{\cal N})^2}.
\label{S}
\ee
The integrated intensity in each of the two peaks are
\be
I(Q\hat x) = \Lambda - I(Q\hat y) = TA_1(\mu-{\cal N}) + \sigma^2A_2(\mu-{\cal N}).
\label{I}
\ee

\begin{figure}
\centering
\subfigure[$\ \sigma/J=0.1, \  b^{eff}/J= \mathbf{0}\ {\rm and}\   0.1$]{
\label{sigma0.1}
\includegraphics[width=8cm]{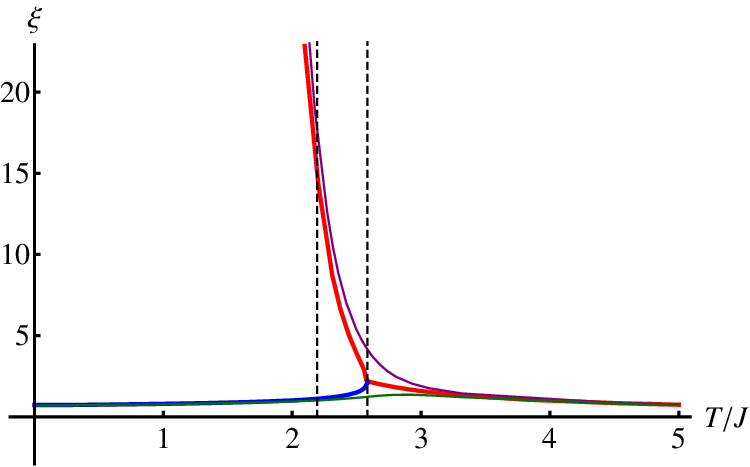}}
\hfill
\subfigure[$\ \sigma/J=1.6, \  b^{eff}/J = \mathbf{0}\ {\rm and}\   0.1$]{
\label{sigma1.6}
\includegraphics[width=8cm]{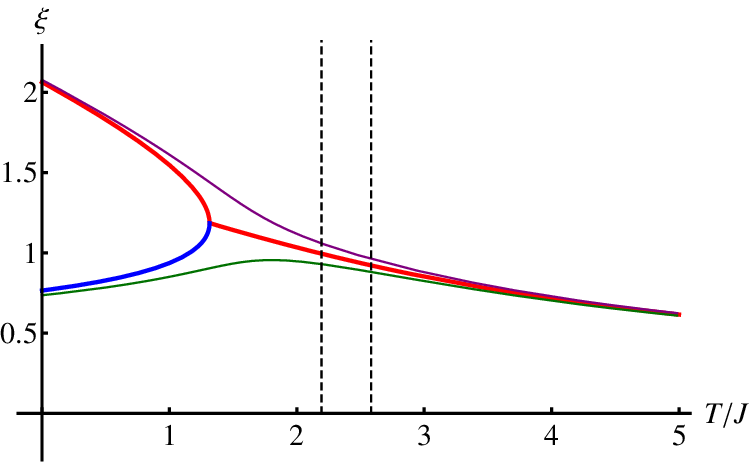}}
\hfill
\subfigure[$\ \sigma/J=\sigma_c/J=2.35, \  b^{eff}/J = \mathbf{0}\ {\rm and}\  0.1$]{
\label{sigmaC}
\includegraphics[width=8cm]{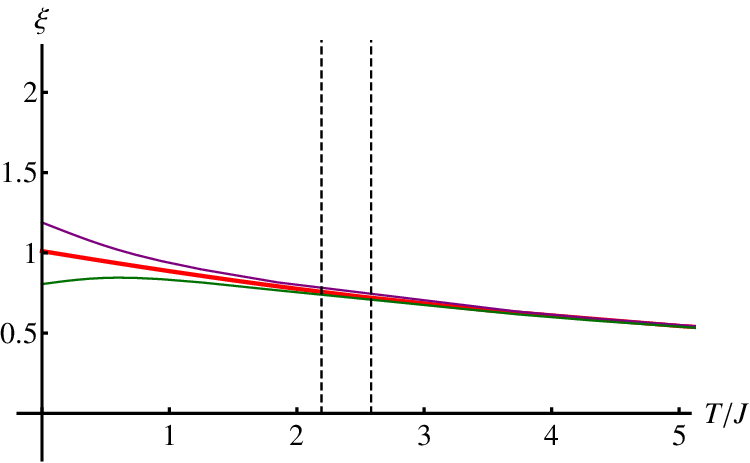}}
%caption v1.2
\caption{$T$-dependence of correlation lengths from the solution of Eqs. (\ref{meanfield}) for various disorder strengths and in the presence or absence of explicit symmetry breaking. Here $J=1, J'=0.01, V_z=0.01, \Delta=0.25,$ and $\xi$ is
the correlation length of $\psi_{\alpha}$ for $\alpha = x$ and $y$ (upper and lower curves, respectively) in units of the lattice constant. (Because we have taken $J'\ll J$, there is little difference between the transverse and longitudinal correlation lengths.)
The clean-limit stripe and nematic transition temperatures are $T_{str} = 2.19J$ and $T_{nem}=2.58J$ respectively, represented by dashed lines in each figures. Thick lines are computed for  $b^{eff}=0$ while for the thin lines there is an explicit symmetry breaking field $b^{eff}=0.1J$.  The critical disorder strength above which there is no nematic transition is $\sigma_c=2.35J$.
Note that the lattice constant of the effective spin model was introduced to provide an ultra-violet  regularization of the theory -- physically, it should be roughly associated with the larger of the CDW period or  the CDW mean-field (amplitude) coherence length.}
\label{correlationlength}
\end{figure}

The nematic character of the state can, in principle, be seen in measures of the CDW structure factor as the difference between properties
near $Q\hat x$ and $Q\hat y$.
Where the CDW correlation length is long, so that $\mu+|{\cal N}| \gg \mu - |{\cal N}|$, this is straightforward.  However, for relatively
short CDW correlation lengths,
where $\mu \gg |{\cal N}|$, the nematic character of the CDW state is relatively subtle.  For instance, from Eq. (\ref{S}),
\be
\frac{S(Q\hat x) -S(Q\hat y) }{S(Q\hat x) +S(Q\hat y) } =\frac {\cal N}{\mu} \left(\frac {T\mu+2\sigma^2}{T\mu + \sigma^2}\right)
+ {\cal O}\left(\frac {\cal N}{\mu}\right)^3.
\ee

In Fig. \ref{sigma1.6}  we exhibit the behavior of the correlation lengths as a function of $T$ for various values of the parameters.  These
were obtained by numerically
solving the saddle-point equations, Eq. (\ref{meanfield}).  It is important to note, before comparing these to experiment, that these were computed assuming a constant (temperature independent) $\Lambda$;  in general, $\Lambda$ (which sets the total amplitude of the CDW correlations) should be only weakly temperature dependent at temperatures small compared to the mean-field $T_c$, but is a strongly decreasing function of $T$ at temperatures approaching the mean-field transition temperature.  Indeed, this effect  enhances the $T$ dependences of all CDW-related correlations at elevated temperatures.

\section{Subtleties and higher order effects}

For the most part, we have focused on the primary order parameters in the problem and have treated explicitly only the lowest order
terms in a Landau-Ginzburg expansion in powers of the order parameter and its gradients.  There are, however, some
subtle pieces of qualitative physics that require higher order terms or that
require analyzing terms beyond saddle-point approximation (or equivalently, terms higher order in powers of $1/N$).
Here we mention a few of these subtleties.

\subsection{Structure at harmonics of the fundamental}

In the absence of disorder, where there is long-range CDW order at one or more of the fundamental ordering vectors, $Q\hat x$, or $Q\hat y$,
one generically expects peaks at harmonics as well, albeit they are generically weaker, as they are higher order in
powers of the order parameter in the regime where the Landau-Ginzburg theory is applicable.  Observation of these harmonics
can be useful in distinguishing the nature of the charge ordered state.
For instance, if there is no way to obtain a macroscopic single-domain order, it may be difficult to distinguish stripe from
checkerboard order by looking only at the fundamentals.  (Sometimes however, due to peculiarities of the crystal
structure, even just looking at the fundamentals may be sufficient to distinguish these two cases, even in the presence of
multiple domains.\cite{robertson2006})

Specifically, in a stripe-ordered state in a tetragonal crystal with an equal number of macroscopic $x$ and $y$ directed
domains, there would be equal strength $\delta$-function peaks in the structure factor at $\mathbf q=Q\hat x$ and $\mathbf q=Q\hat y$,
just as there would be for a checkerboard ordered state.  However,
while both states would also exhibit weaker second harmonic peaks at  $\mathbf q=2Q\hat x$ and $\mathbf q=2Q\hat y$, the checkerboard state
would also exhibit a second-harmonic peak at  $\mathbf q=\mathbf Q_{xy}\equiv Q\hat x+Q\hat y$ which would be absent in the
stripe-ordered state.  So it is reasonable to ask whether the
same is true of the not-quite-ordered CDW state in the presence of non-zero disorder.

The  structure factor in the neighborhood of these second harmonics is the Fourier transform of the correlation functions of the bilinear
order parameter, $\tilde S_{\alpha\alpha^\prime}(\mathbf r)$.  In the $U(1)$ representation, where $\psi_\alpha$ is a complex
scalar field, the second harmonic is also a
complex scalar field, $\psi_{\alpha\alpha^\prime}=\psi_\alpha\psi_\alpha'$ which transforms under translations as
$\psi_{\alpha\alpha^\prime}\to e^{iQ(r_\alpha+r_{\alpha^\prime})}\psi_{\alpha\alpha^\prime}$.  In the $SO(2)$ representation,  where
$\psi_{i\alpha}$ is a real vector field with $\psi_{1\alpha}={\rm Re}[\psi_{\alpha}]$ and $\psi_{2\alpha}={\rm Im}[\psi_{\alpha}]$, the same composition
law (written in a way that is straightforward to generalize to $SO(N)$  is (adopting summation convention)
\be
\label{second}
\psi_{\alpha\alpha^\prime;jj^\prime}(\vec r,m)=\frac {g_{2}} N\psi_{\alpha,i}(\vec r,m)\Gamma^{(jj^\prime)}_{i,i^\prime}
\psi_{\alpha^\prime,i^\prime}(\vec r,m)
%&&\psi_{2y,jj^\prime}(\vec r,m)=g_{2}\psi_{y,i}(\vec r,m)\Gamma^{(jj^\prime)}_{i,i^\prime}\psi_{y,i^\prime}(\vec r,m) \nonumber\\
%&&\psi_{xy,jj^\prime}(\vec r,m)=g_{2}\psi_{x,i}(\vec r,m)\Gamma^{(jj^\prime)}_{i,i^\prime}\psi_{y,i^\prime}(\vec r,m),\nonumber
\ee
where $\Gamma$ are the traceless symmetric tensors,
\be
\Gamma^{(ij)}_{kp} = \Gamma^{(ij)}_{pk} = \Gamma^{(ji)}_{pk}=\delta_{ik}\delta_{jp}+ \delta_{ip}\delta_{jk} - \frac 2 N \delta_{ij}\delta_{pk},
\ee
such that
\be
\Gamma^{(ij)}_{pk} \Gamma^{(ij)}_{p^\prime k^\prime} = 2\Gamma^{(pk)}_{p^\prime k^\prime}\ .
\ee
In terms of these,
\be
\tilde S_{xy}(\vec r,m)=  \sum_{ij}\overline{\big\langle \psi_{xy;ij}(\vec r,m)\psi_{xy;ij}(\vec 0,0)\big\rangle}.
\ee
and similarly for $\tilde S_{xx}$ and $\tilde S_{yy}$.

To lowest order in $1/N$, this means that the second harmonic structure factor is simply a convolution of the primaries, as in Eq. (12) in the main text.
This expression has no explicit dependence on $\Delta$, and so does not depend any more sensitively than do the  fundamentals on the sign of $\Delta$
(which would determine whether stripes or checkerboards were favored in the absence of disorder).  The first correction that brings in an explicit dependence on $\Delta$ gives
 \bea
 S_{\alpha\alpha^\prime}(\mathbf k)&& =2(g_2)^2\left[1+\delta_{\alpha\alpha^\prime}-\frac {2\delta_{\alpha\alpha^\prime}}N\right]
%\int \frac {d^3 q}{(2\pi)^3} S_\alpha(\mathbf k+\mathbf q)S_{\alpha^\prime}(-\mathbf q)
 \Pi_{\alpha\alpha'}(\mathbf k)\\
 && - \frac {4(g_2)^2} N \int \frac {d^3 q}{(2\pi)^3}
 \frac {d^3 q^\prime}{(2\pi)^3} S_\alpha(\mathbf k+\mathbf q)S_{\alpha^\prime}(\mathbf q)
 V_{\alpha\alpha^\prime}(\mathbf q-\mathbf q^\prime)S_\alpha(\mathbf k+\mathbf q^\prime)S_{\alpha^\prime}(\mathbf q^\prime)  \nonumber \\
&& +{\cal O}\left( N^{-2}\right)\nonumber
 \eea
 where
 \bea
&& V_{xy}(\mathbf k) =2D(\mathbf k)^{-1}\big[ U+\Delta\big]   \\
&& V_{xx}(\mathbf k) =2D(\mathbf k)^{-1}\big[(U-\Delta) -4U\Delta \Pi_{yy}(\mathbf k)\big]
\nonumber \\
&& V_{yy}(\mathbf k)=2 D(\mathbf k)^{-1}\big[(U-\Delta) -4U\Delta \Pi_{xx}(\mathbf k)\big]
\nonumber \\
&&D(\mathbf k)=1 +2(U-\Delta)[\Pi_{xx}(\mathbf k) + \Pi_{yy}(\mathbf k)] - 8U \Delta \Pi_{xx}(\mathbf k)  \Pi_{yy}(\mathbf k)
\nonumber
\eea
and where
\be
\Pi_{\alpha\alpha'}(\mathbf k)=\int \frac{d^3 q}{(2\pi)^3} S_{\alpha}(\mathbf q)S_{\alpha'}(\mathbf k+ \mathbf q).
\ee
This is a complicated expression, but the qualitative point can be seen directly:  The leading order term contains
no additional information to distinguish stripe and checkerboard orders that is not already apparent in the structure factor near the
fundamental ordering vectors.  The first $1/N$ correction is
generally negative, {\it i.e.} it tends to suppress the magnitude of the harmonic peaks, but it does depend explicitly on the sign of $\Delta$.
In particular, for positive $\Delta$, the structure at $2Q\hat x$ and $2Q\hat y$ are supressed less than the structure at $\mathbf Q_{xy}$,
while negative $\Delta$ has the opposite effect.

The expression can be somewhat simplified in the hard spin limit $U\to\infty$, where
\bea
&& V_{xy} \to \big[\Pi_{xx}+\Pi_{yy}-4\Delta \Pi_{xx}\Pi_{yy}\big]^{-1} \\
&& V_{xx} \to V_{xy}\big[1-4\Delta \Pi_{yy}\big]
\nonumber \\
&& V_{yy} \to V_{xy}\big[1-4\Delta \Pi_{xx}\big] .
\nonumber
\eea
Here the qualitative response to the sign of $\Delta$ is apparent.  However, it is clear that unless there is a very pronounced peak at the harmonic ordering
vector (so that $\Delta S_{\alpha\alpha}$ is significant), such effects will be subtle and difficult to interpret.
  \subsection{Shifts of the ordering vector}
 One unphysical feature of the model we have treated is that the ordering wave vector is constant, independent of $T$ and $\sigma$ and any of the other
 variables.  In contrast, incommensurate density waves generically have $T$ dependent ordering vectors.  This can be corrected by including higher order
 terms in the effective field theory - of which the lowest order terms are
\bea
\label{otherterms}
\delta {\cal H} =&&
 \frac {g_+} {iN} \Big[\big|\psi_{x}\big|^2+\big|\psi_{y}\big|^2-N\Lambda\Big]\Big[ \psi^\dagger_{x}\partial_x\psi_{x}
 +\psi^\dagger_{y}\partial_y\psi_{y}\Big] \\
+&& \frac {g_-} {iN} \Big[\big|\psi_{x}\big|^2-\big|\psi_{y}\big|^2\Big]\Big[ \psi^\dagger_{x}\partial_x\psi_{x}-\psi^\dagger_{y}\partial_y\psi_{y}\Big] +
\ldots .
 \nonumber
\eea
At first blush, these terms appear to violate inversion symmetry, but it is important to recall that zero momentum
in the effective field theory actually corresponds to
momentum $Q\hat e_\alpha$ in physical terms.  Thus, positive momenta add to $Q$ while negative momenta reduce it in magnitude.

The first term here produces a generally $T$ and $\sigma$ dependent shift in the magnitude of the ordering vector, but it vanishes in the hard-spin limit.
There still may be some smooth $T$ dependence of $Q$ which comes from high energy physics and which appears as an analytic
temperature dependence of $Q$ that can be
included explicitly, but which does not reflect any of the emergent physics of a growing CDW correlation length.

The second term is significant in the nematic phase, where it produces a relative shift between the ordering vector in the $x$ and $y$,
which to leading order in $1/N$ is
\be
\delta \mathbf Q_x =  \left(\frac {\cal N}{2\Delta\kappa_{\|}}\right)  \hat x \ \ {\rm and} \ \ \delta \mathbf Q_y =
- \left(\frac {\cal N}{2\Delta\kappa_{\|}}\right)  \hat y .
\ee

\subsection{Coupling to strain}
A major difference between ${\bf Q}={\bf 0}$ and non-zero orders is the implications of their coupling to strain:  For non-zero ${\bf Q}$, the induced interactions fall exponentially with distance, and so if the coupling to the lattice is weak, the effects are negligible.  By contrast, for ${\bf Q}={\bf 0}$ order, including nematicity, strain-induced effective interactions are long-ranged, and hence can have important consequences even if weak.  One particularly important consequence of this is that even if the electronic structure is quasi-2D ({\it i.e.} $V_z \ll J$), so that the CDW correlations are essentially confined to single planes, the interplane nematic couplings can  none-the-less  be significant.  Such strain effects first appear in the effective field theory through terms of the form
\be
\delta {\cal H} =\ldots +  {g_{strain}}\big[\epsilon_{xx} - \epsilon_{yy}\big] \big[\psi^\dagger_{x}\psi_x - \psi^\dagger_y\psi_y\big] +\ldots
\ee
where $\epsilon_{\alpha\alpha^\prime}$ is the strain tensor.  Not coincidently, this term also embodies the coupling of the nematic order to any small orthorhombicity of the crystal, where in this case $\epsilon_{\alpha\alpha^\prime}$ is the orthorhombic strain defined relative to a putative tetragonal parent compound.

 \section{Are the CDW sightings in the different hole doped cuprates closely related?}

%The CDW correlations in the various families of underdoped cuprates differ in some details, but are otherwise very similar.  In all cases, the CDW structure
%factor as inferred from  X-ray or neutron scattering or from STM, is peaked at two pairs of  incommensurate ``CDW ordering vectors" $\vec Q=\pm (2\pi/\lambda)\hat x$ and
%$\vec Q=\pm (2\pi/\lambda)\hat y$ where, depending on the material and the doped hole concentration, $x$, the CDW period $\lambda$ is 3 or 4 lattice constants
%or sometimes even longer,  the peak widths are not resolution limited, but rather correspond to CDW correlation lengths in the range from about 1.5 $\lambda$ in
%BSCCO, thru 6-8 $\lambda$ in YBCO, up to 25 $\lambda$ in LBCO, and always have extremely short correlation lengths (of order the spacing between planes) in the $z$ direction.

There has been some debate about whether  the CDW tendencies seen in the various different cuprates are close siblings or
many-times removed cousins - {\it i.e.}
whether the differences from one family of cuprates to another are the expected ``small'' effects produced by the somewhat
different crystalline environment and degree
of quenched disorder  in the different materials, or are so ``large'' that  they should be thought of as different phenomena with
different mechanisms.  This latter viewpoint
seems untenable to us, for reasons that are elaborated elsewhere.\cite{fradkin2012,neto2013}

It is, however, worth mentioning that there is very compelling evidence from transport that the basic charge-ordering phenomena are
extremely closely related in all the
families of hole-doped cuprates.  Specifically, several transport signatures of the incipient charge order have been identified by the
group of Taillefer\cite{Taillefer2012,Chang2010,laliberte2011,doironleyrad2013} by studying various stripe-ordered 214 materials,
including LBCO, NdLSCO, and EuLSCO.
Because the CDW order has particularly long correlation lengths in these materials (and hence is easier to identify in scattering experiments),
they were able to correlate the diffraction data with salient features of the transport data.  This identification is significant in its own
right - it shows that the CDW ordering phenomena have a significant effect on the low energy itinerant electronic structure,
{\it i.e.} that it is an ``important'' actor in the electronic physics of these materials.

The Taillefer group has then measured the same transport properties in YBCO and Hg1201 in the same range of copings and seen
extraordinarily similar features.  In some cases, transport data\cite{laliberte2011,doironleyrad2013} from NdLSCO, YBCO, and
Hg1201 at the same doping can  be lain on top of each other and are
essentially indistinguishable.  (The CDW transition in LBCO is sharper than in the other materials, as reflected in its longer
correlation length, and correspondingly the associated features in the transport are anomalously sharp in this material.)  It is
difficult to imagine that there could be significant differences in the nature of the
charge ordering in the different families of hole doped cuprates, given the great similarities between the transport signatures.

\bibliography{DisorderCDW}